\documentclass{jpp}
\usepackage{graphicx}
\usepackage{hyperref}
\usepackage{epstopdf, epsfig}
\usepackage{layouts}
\usepackage{subcaption}

\usepackage[utf8]{inputenc}
\usepackage[T1]{fontenc}
\usepackage{amsmath}
\usepackage{lipsum}

\shorttitle{Available energy in tokamak equilibria}
\shortauthor{Mackenbach et al.}

\title{Available energy of trapped electrons in Miller tokamak equilibria}

\author{R.J.J. Mackenbach\aff{1,2}
  \corresp{\email{r.j.j.mackenbach@tue.nl}},
  J.H.E. Proll\aff{1},
  G. Snoep\aff{1,3},
 \and P. Helander\aff{2}}

\affiliation{\aff{1}Eindhoven University of Technology, 5612 AZ Eindhoven, The Netherlands
\aff{2}Max Planck Institute for Plasma Physics, 17491 Greifswald, Germany
\aff{3}DIFFER - Dutch Institute for Fundamental Energy Research, Eindhoven, The Netherlands}

\begin{document}

\maketitle

\begin{abstract}
Available energy (\AE{}), which quantifies the maximum amount of thermal energy that may be liberated and converted into instabilities and turbulence, has shown to be a useful metric for predicting saturated energy fluxes in trapped-electron-mode-driven turbulence. Here, we calculate and investigate the \AE{} in the analytical tokamak equilibria introduced by \citet{Miller1998NoncircularModel}. The \AE{} of trapped electrons reproduces various trends also observed in experiments; negative shear, increasing Shafranov shift, vertical elongation, and negative triangularity can all be stabilising, as indicated by a reduction in \AE{}, although it is strongly dependent on the chosen equilibrium. Comparing \AE{} with saturated energy flux estimates from the \textsc{tglf} model, we find fairly good correspondence, showcasing that \AE{} can be useful to predict trends. We go on to investigate \AE{} and find that negative triangularity is especially beneficial in vertically elongated configurations with positive shear or low gradients. We furthermore extract a gradient threshold-like quantity from \AE{} and find that it behaves similarly to gyrokinetic gradient thresholds: it tends to increase linearly with magnetic shear, and negative triangularity leads to an especially high threshold. We next optimise the device geometry for minimal \AE{} and find that the optimum is strongly dependent on equilibrium parameters, e.g. magnetic shear or pressure gradient. Investigating the competing effects of increasing the density gradient, the pressure gradient, and decreasing the shear, we find regimes that have steep gradients yet low \AE{}, and that such a regime is inaccessible in negative-triangularity tokamaks.
\end{abstract}

\section{Introduction} \label{sec:outline}
Energy transport in tokamaks and stellarators is largely dominated by turbulent energy losses, which severely degrade the energy confinement in these devices. A detailed understanding of how various parameters characterising the plasma and the magnetic field geometry, such as magnetic shear and the pressure gradient, affect the turbulent transport properties would be helpful in comprehending and mitigating this. The standard method to assess the turbulence properties of any given tokamak is to perform nonlinear gyrokinetic simulations. However, such simulations are computationally expensive because of the very disparate time- and length scales characterising the turbulence and the transport. Thus, it would be beneficial to find a reduced model capable of predicting the level of turbulent transport by simpler means.
\par In a recent publication, it was shown that the available energy (\AE{}) of trapped electrons can serve as such a reduced model \citep{Mackenbach2022AvailableTransport}, at least for turbulence driven by the plasma density gradient. Any plasma possesses a maximum amount of thermal energy that can be converted into instabilities and turbulence \citep{Gardner1963BoundPlasma}. This ``available'' energy can be calculated by performing a Gardner restacking of the plasma distribution function $f$, in which phase-space volume elements are rearranged in a manner that respects Liouville's theorem \citep{Kolmes2020AvailableRearrangements,Kolmes2020RecoveringOperations}. The restacking of $f$ that minimises the thermal energy results in a ``ground state'' distribution function $f_g$, and the \AE{} is defined as the difference in thermal energy between $f$ and $f_g$. If one imposes the additional constraint that adiabatic invariants be conserved in the restacking process, the \AE{} becomes relevant to magnetically confined plasmas \citep{Helander2017AvailablePlasmas,Helander2020AvailablePlasmas}. In fusion plasmas, the magnetic moment $\mu$ is generally conserved for all species, and the parallel adiabatic invariant $\mathcal{J} = \int m v_\| \: \mathrm{d} \ell$ is conserved for magnetically trapped electrons. \par 
A significant portion of the electrons are trapped and can contribute to turbulence through trapped electron modes (TEMs). The \AE{} of trapped electrons correlates with the turbulent energy flux for such TEM-driven turbulence over several orders of magnitude in saturated energy fluxes \citep{Mackenbach2022AvailableTransport}. This correlation is expressible as a simple power law, where the saturated energy flux, $Q_\text{sat}$, was found to be related to the available energy, which we denote by $A$ in formulas, via approximately
\begin{equation}
    Q_\text{sat} \propto A^{3/2}.
\end{equation}
This relation was found to hold for both a tokamak and stellarators, and for various values of the density gradient. Aside from this relationship, other links have been found by \cite{kolmes2022minimum} where quasi-linear plateauing is shown to be related to a concept closely connected to \AE{}, highlighting other links to transport physics. In any case, to gain a deeper understanding, it is of interest to derive an explicit expression of \AE{} in tokamak geometry, in order to investigate the dependence of it on various geometrical and plasma parameters. \par 
This is our aim in the present paper, where we compute \AE{} for the family of tokamak equilibria constructed by \citet{Miller1998NoncircularModel}. The starting point is the following explicit expression for \AE{} in a flux tube of any omnigenous equilbrium \citep{Helander2020AvailablePlasmas,Mackenbach2023AvailableTransport}, including that of a tokamak,
\begin{equation}
    A= \frac{1}{2 \sqrt{\pi}} \frac{\upi L \Delta \psi_t  \Delta \alpha_C }{B_0} \iint \sum_{{\rm wells}(\lambda)} e^{-z} z^{5/2} \hat{\omega}_{\alpha}^{2} \mathcal{R}\left[ \frac{1}{z} \frac{\hat{\omega}_{*}^{T}}{\hat{\omega}_{\alpha}}-1\right] \hat{g}^{1/2} \mathrm{d} \lambda \mathrm{d}z.
    \label{eq:AE-of-fluxsurface-general}
\end{equation}
Here, $L$ is the total length of a field-line completing one poloidal turn, $B_0$ is some reference magnetic field strength, $z=H/T_0$ is the particle energy normalised by the temperature, $\lambda = \mu B_0/H$ is the pitch angle, and $\Delta \psi_t$ and $\Delta \alpha_C$ denote the size of the flux-tube in the radial and binormal directions respectively (we have parameterised the radial coordinate by means of the toroidal flux $\psi_t$ and the binormal by means of the Clebsch angle $\alpha_C$). We furthermore sum over all magnetic wells with a certain value $\lambda$. The hatted quantities in the integrand denote normalised frequencies, with $\hat{\omega}_\alpha$ being the normalised bounce-averaged drift precession frequency, $\hat{\omega}_*^T$ the normalised electron diamagnetic drift frequency, and $\hat{g}^{1/2}$ the normalised bounce time. They are explicitly defined as
\begin{subequations}
\begin{align}
    \hat{\omega}_\alpha & \equiv - \frac{\Delta \psi_t}{H} \frac{\partial_{\psi_t} \mathcal{J}}{\partial_{H} \mathcal{J}}, \label{eq:precession-freq} \\
    \hat{\omega}_*^T & \equiv \Delta \psi_t \frac{\mathrm{d} \ln n}{\mathrm{d} \psi_t} \left( 1 + \eta \left[ z - \frac{3}{2} \right] \right), \label{eq:diamagnetic-drift} \\
    \hat{g}^{1/2} & \equiv \frac{\partial_H\mathcal{J}}{L} \sqrt{\frac{2H}{m}}, \label{eq:normalised-bouncetime}
\end{align}
\end{subequations}
where we have denoted the ratio between the gradients by $\eta = (\mathrm{d} \ln{T} / \mathrm{d} \psi_t ) / (\mathrm{d} \ln{n} / \mathrm{d} \psi_t )$. Finally, $\mathcal{R}[x]=(x+|x|)/2$ is the ramp function. Using the above expressions, we shall find the \AE{} of trapped electrons in any Miller tokamak.

\section{Theory}

\subsection{The available energy in any omnigenous system}
\begin{figure}
    \centering
    \includegraphics[width=0.5\textwidth]{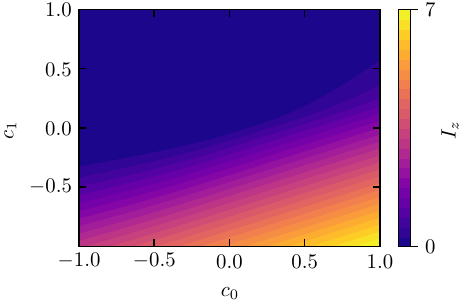}
    \caption{Contour plot of $I_z$ as a function of $c_0$ and $c_1$.}
    \label{fig:I_z-plot}
\end{figure}

We first note that the integral over $z$ can be rewritten into a convenient form. We define two functions that are independent of $z$, namely,
\begin{equation}
    c_0 = \frac{ \Delta \psi_t}{\hat{\omega}_\alpha(\lambda)} \frac{\mathrm{d} \ln(n)}{\mathrm{d} \psi_t} \left( 1 - \frac{3}{2} \eta \right), \qquad c_1 = 1 - \frac{\Delta \psi_t}{\hat{\omega}_\alpha(\lambda)} \frac{\mathrm{d} \ln(n)}{\mathrm{d} \psi_t} \eta.
    \label{eq: c0 and c1}
\end{equation}
With these functions, the integral over the normalised energy $z$ reduces to the following form;
\begin{equation}
    I_z(c_0,c_1) = \frac{8}{3 \sqrt{\upi}} \int_0^\infty \exp(-z) z^{3/2} \mathcal{R} \left[ c_0 - c_1 z \right] \mathrm{d} z.
\end{equation}
This integral can be solved analytically, and its functional form depends on the signs of $c_0$ and $c_1$, resulting in four different conditions. The easiest case to evaluate is the case where $c_0<0$ and $c_1>0$. In this case, the argument of the ramp function is always negative, and hence the integral reduces to zero. The second case is when the argument of the ramp function is always positive, which occurs whenever $c_0\geq0$ and $c_1 \leq 0$. The integral then reduces to the following form,
\begin{equation}
    I_z = 2c_0 - 5c_1 .
\end{equation}
There are two cases left to consider. First, we inspect the case where the argument of the ramp function is positive for low $z$ but becomes negative for high $z$, that is, $c_0 \geq 0$ and $c_1 > 0$. The unique point where the argument of the ramp function vanishes is the following,
\begin{equation}
    z_* = \frac{c_0}{c_1}.
\end{equation}
Thus, the integral becomes
\begin{equation}
    I_z = \frac{8}{3 \sqrt{\upi}} \int_0^{z_*} \exp(-z) z^{3/2} \left( c_0 - c_1 z \right) \mathrm{d} z.
\end{equation}
This integral can be expressed in terms of the error function, $\text{erf}(x) = 2/\sqrt{\upi} \int_0^x \exp(-t^2) \mathrm{d}t$,
\begin{equation}
    I_z =  (2c_0 - 5c_1) \text{erf}\left({\sqrt{\frac{c_0}{c_1}}}\right) + \frac{2}{3\sqrt{\upi}} (4c_0 + 15c_1) \sqrt{\frac{c_0}{c_1}} \exp\left( - \frac{c_0}{c_1} \right).
\end{equation}
The final case is that where the argument of the ramp function is negative for low $z$ but becomes positive for high $z$, that is, $c_0 < 0$ and $c_1 \leq 0$. The integral then becomes
\begin{equation}
    I_z =(2c_0 - 5c_1) \left[ 1 - \text{erf}\left({\sqrt{\frac{c_0}{c_1}}}\right) \right] -\frac{2}{3\sqrt{\upi}} (4c_0 + 15c_1) \sqrt{\frac{c_0}{c_1}} \exp\left( -\frac{c_0}{c_1} \right).
\end{equation}
Note that $I_z \geq 0, \; \forall(c_0,c_1) \in \mathbb{R}^2$, which can also be seen in Fig. \ref{fig:I_z-plot}.
The \AE{} can now be found by executing the integral over the remaining coordinate
\begin{equation}
    A = \frac{3}{16} \frac{\Delta \psi_t \Delta \alpha_C L}{B_0} n_0 T_0 \int_{\{ \lambda \}} \mathrm{d} \lambda  \sum_{\text{wells}(\lambda)}   I_z(c_0,c_1) \hat{\omega}_\alpha^2 \hat{g}^{1/2}.
    \label{eq:AE-new-coords}
\end{equation}
Note that this expression is completely general; no approximations have been made in executing these integrals, aside from the preceding assumption of omnigeneity. \par 
It is also interesting to note that from this expression one can see that there are no tokamak configurations with vanishing \AE{}, at least in leading order near the axis. This conclusion can most readily be drawn by investigating the expression for $\omega_\alpha$ from \citet{Connor1983EffectTokamak}. Here, one can find that there is always a zero crossing for $\omega_\alpha$ (with no pressure gradient), which implies that $c_0$ and $c_1$ must change sign. As such, the available energy \emph{must} be non-zero (as either $I(c_0,c_1)$ or $I(-c_0,-c_1)$ must be non-zero). Formally, this corresponds to the fact that such a zero crossing implies that the device does not have the so-called maximum-$\mathcal{J}$ property, which is required for the linear stability of trapped electron modes \citep{proll2012resilience}.\footnote{This correspondence between the maximum-$\mathcal{J}$ property and \AE{} is shown in \citep{Helander2017AvailablePlasmas}, and can also be understood from Eq. \eqref{eq:AE-new-coords}. A device is said to be maximum-$\mathcal{J}$ if $\partial_\psi \mathcal{J}<0$ for all particles, which implies
$\hat{\omega}_\alpha > 0$ for all $\lambda$.  For $\eta < 2/3$, Eq. \eqref{eq: c0 and c1} implies that $c_0 < 0$ for a radially decreasing density profile, and $c_1 > 0$, thus the integrand of the \AE{} reduces to zero since $I_z = 0$.} \par 
To make further progress in solving Eq. \eqref{eq:AE-new-coords}, one requires the function $\hat{\omega}_\alpha(\lambda)$, which in turn requires a specification of the equilibrium. In this paper, we will use the local construction of the equilibrium, employing a formalism developed by \citet{MercierLuc1974}.

\subsection{Construction of local equilibria}
Equilibria are constructed by finding a radially local solution to the Grad-Shafranov equation, and this solution allows us to find $\hat{\omega}_\alpha$. We highlight the essential components of this derivation, which essentially follows the steps taken by \citet{Miller1998NoncircularModel}, and a thorough overview is given by \citet{candy2009unified} \par 
The Mercier-Luc formalism requires the shape of the flux surface, the poloidal field $B_p$ on that flux surface, the gradients of the pressure $p(\psi)$, and the toroidal field function $f(\psi) = R B_\phi$ on the flux surface, where $R$ is the major radial coordinate, $B_\phi$ is the toroidal component of the magnetic field, and $\psi$ is the poloidal flux. We parameterize the flux surface as $R_s = R_s(l)$ and $Z_s = Z_s(l)$, where $l$ measures the poloidal arclength along the flux surface. It is also useful to define a tangential angle $u$, which measures the angle between the unit vector in the major radial direction $\boldsymbol{e}_R$ and the vector tangential to the flux surface $\boldsymbol{e}_l$ clockwise, thus
\begin{subequations}
\begin{eqnarray}
\frac{\mathrm{d}R_s(l)}{\mathrm{d} l} &=& \cos u,
\\
\frac{\mathrm{d}Z_s(l)}{\mathrm{d} l} &=& - \sin u. 
\end{eqnarray}
\end{subequations}
With this definition, the angle $u$ can be calculated by $\mathrm{d} u / \mathrm{d} l = - 1 / R_c$, where $R_c(l)$ is the radius of curvature of the poloidal cross section, and the negative sign arises because the poloidal arclength is measured clockwise. We go on to introduce a radial-like expansion variable $\rho$ which is zero on the given flux surface, in terms of which the cylindrical coordinates become
\begin{subequations}
\begin{eqnarray}
R(\rho,l) &=& R_s(l) + \rho \sin u, 
\\
Z(\rho,l) &=& Z_s(l) + \rho \cos u. 
\end{eqnarray}
\end{subequations}
The metric tensor in these coordinates has non-zero components only on the diagonal (which is to be expected as we ensured orthogonality in the construction),
\begin{equation}
    g_{ij} = \text{diag}\left[ \left(1 - \frac{\rho}{R_c} \right)^2, \; 1, \; R^2 \right],
\end{equation}
where we use the convention $x^1 = l$, $x^2 = \rho$, $x^3 = \phi$. The local solution is now constructed by expanding in $\rho$, 
\begin{subequations}
\begin{eqnarray}
\psi &\approx& \psi_0 + \rho \psi_1 + \frac{\rho^2 }{2}\psi_2, \\
p'(\psi)    &\approx& p'(\psi_0)  \\
f'(\psi)    &\approx& f'(\psi_0)  
\end{eqnarray}
\end{subequations}
and substitute into the Grad-Shafranov equation, which in leading order reduces to
\begin{equation}
    \psi_2 = \left( \sin(u) + \frac{R_s}{R_c} \right) \frac{\psi_1}{R_s} - \mu_0 R_s^2 p'(\psi_0) - f(\psi_0)f'(\psi_0).
\end{equation}
This allows one to find the radial variation of the poloidal magnetic field by using \citep{Helander2005CollisionalPlasmas}
\begin{equation}
    B_p = \frac{ | \nabla \psi |}{R},
    \label{eq:pol-magfield}
\end{equation}
resulting in
\begin{equation}
    B_p(l,\rho) = \frac{\psi_1}{R_s} \left( 1 + \rho  \left[ \frac{1}{R_c} - \frac{ \mu_0 R_s^2 p'(\psi_0)}{\psi_1} - \frac{f(\psi_0) f'(\psi_0)}{\psi_1} \right] \right).
    \label{eq:Bp}
\end{equation}
From this equation we can immediately see that $\psi_1/R_s = B_{p,s}$, with $B_{p,s}$ being the poloidal field on the flux-surface as indicated by the subscript. As such, the poloidal field strength can be written as
\begin{equation}
   \left| B_p(l,\rho)\right| = B_{p,s}\left( 1 + \rho \left[ \frac{1}{R_c} - \frac{\mu_0 R_s p'}{B_{p,s}} - \frac{ff'}{R_s B_{p,s}} \right] \right) \equiv B_{p,s} \left( 1 + \rho \p_\rho b_p \right)
\end{equation}
The toroidal field is found from its definition $B_\phi = f(\psi)/R$, resulting in
\begin{equation}
    \left|B_\phi(l,\rho) \right| = B_{\phi,s} \left( 1 + \rho \left[ \frac{ f'(\psi_0)}{f(\psi_0)} R_s B_{p,s} - \frac{\sin u}{R_s} \right] \right) \equiv B_{\phi,s} \left( 1 + \rho \p_\rho b_\phi \right),
    \label{eq:Bphi}
\end{equation}
where $B_{\phi,s} = f(\psi_0)/R_s$. The total magnetic field strength is also readily derived
\begin{equation}
    B = \sqrt{B_{\phi,s}^2+B_{p,s}^2} \left(1 + \rho \left[ \frac{B_{\phi,s}^2 \p_\rho b_\phi +  B_{p,s}^2 \p_\rho b_p }{B_{\phi,s}^2+B_{p,s}^2} \right] \right) \equiv B_s \left( 1 + \rho \p_\rho b \right).
\end{equation}
Note that the derivatives $\partial_\rho b_p$, $\partial_\rho b_\phi$, and $\partial_\rho b$ are given in square brackets. The radial variation of the poloidal line element is readily found from the metric tensor, 
\begin{equation}
    \mathrm{d}l = \left(1 - \frac{\rho}{R_c}\right) \left[ \mathrm{d}l \right]_{\rho=0}
\end{equation}
In these equations $f'(\psi_0)$ is treated as a free parameter, but it is difficult to ascertain if the chosen value of this parameter is realistic. It is more convenient, however, to specify the magnetic shear, which is related to $f'(\psi_0)$. This can be made explicit by investigating the safety factor
\begin{equation}
    q = \frac{f(\psi)}{2 \upi} \int \frac{ \mathrm{d}l }{R_s^2 B_{p,s}}.
    \label{eq:safety-factor}
\end{equation}
Taking the derivative of the safety factor with respect to $\psi$, one finds an equation describing this relationship,
\begin{equation}
    \p_\psi q = \frac{f'}{f} q + f \frac{1}{2\pi} \int \frac{\mathrm{d}l}{R_s^3 B_{p,s}^2} \left( - \frac{2}{R_c} - \frac{2 \sin u}{R_s} + \frac{\mu_0 R_s p'}{B_{p,s}} + \frac{f f'}{R_s B_{p,s}} \right).
    \label{eq:relationship-shear-fp}
\end{equation}
We also wish to relate the arclength along a magnetic field line to the poloidal arclength. These quantities are related as
\begin{equation}
    \mathrm{d}\ell =\left| \frac{B}{B_p} \right| \mathrm{d}l.
\end{equation}
Finally, the poloidal coordinate can be expressed in terms of the poloidal angle $\theta$ rather than the poloidal arclength by
\begin{equation}
    l_\theta \equiv \frac{\mathrm{d}l}{\mathrm{d}\theta} = \sqrt{ (\p_\theta R_s)^2 + (\p_\theta Z_s)^2 },
\end{equation}
and the total arclength thus becomes 
\begin{equation}
    L = \oint l_\theta \left| \frac{B}{B_p} \right| \mathrm{d}\theta.
\end{equation}

\subsection{Non-dimensionalisation and available energy}
We proceed to make the various functions dimensionless as in \citet{Roach1995TrappedTokamaks}, and in doing so we will introduce various dimensionless constants which will be useful for the remainder of the analysis. We assume that we have been given the dependencies of the various functions in terms of the minor radial coordinate $r$, which in turn relates to the major radial coordinate $R_0$ through the inverse aspect ratio of the flux surface in question $\epsilon=r/R_0$. Furthermore, we define our reference field $B_0$ through the relation $f(\psi_0) = B_0 R_0$. Let us now define various dimensionless functions of interest,
\begin{subequations}
\begin{eqnarray}
\hat{R}_s       &=& R/R_0,  \\
\hat{Z}_s       &=& Z/R_0,  \\
\hat{R}_c       &=& R_c/r,  \\
\hat{l}_\theta  &=& l_\theta/r,  \\
\hat{B}_\phi    &=& B_\phi/B_0,  \\
\hat{B}         &=& B/B_0. 
\end{eqnarray}
\end{subequations}
One also needs to relate $\psi$ to $r$, which can be done by investigating the poloidal field as in Eq. \eqref{eq:pol-magfield}
\begin{equation}
    B_p = \frac{\p_r \psi}{R_0} \frac{ \left| \nabla r \right|}{\hat{R}_s},
    \label{eq:poloidal-field}
\end{equation}
We go on to identify two factors in the above expression, namely
\begin{equation}
    B_{p,0} \equiv \p_r \psi /R_0 
\end{equation} 
and 
\begin{equation}
    \hat{B}_{p,s} \equiv \left| \nabla r \right| / \hat{R}_s.
\end{equation} Inserting these into the equation for the safety factor $\eqref{eq:safety-factor}$, one finds
\begin{equation}
    B_{p,0} = \frac{\gamma \epsilon}{q} B_0, \quad \gamma \equiv \frac{1}{2 \upi} \oint \frac{\hat{l}_\theta}{\hat{R}_s^2 \hat{B}_{p,s}} \mathrm{d} \theta.
    \label{eq: gamma def}
\end{equation}
We proceed to define a dimensionless pressure gradient, analogous to the $\alpha$ parameter used in $s$-$\alpha$ geometry,
\begin{equation}
    \alpha = - \frac{2 \mu_0 \epsilon^2 R_0^2 p'}{B_{p,0}} = -\epsilon r \frac{\mathrm{d} p}{\mathrm{d} r} \Bigg/ \frac{B_{p,0}^2}{2 \mu_0}.
\end{equation}
Note that this dimensionless pressure gradient is not related to the Clebsch angle. The pressure gradient can in turn be used to define a dimensionless toroidal current density
\begin{equation}
    \sigma = \left( \mu_0 p' + \frac{ff'}{R_0^2} \right) \frac{\epsilon R_0^2}{B_{p,0}} = \frac{q}{\gamma} f' R_0 - \frac{\alpha}{2 \epsilon}.
\end{equation}
We go on to define the shear $s$ in the following manner
\begin{equation}
    s = \epsilon R_0^2 B_{p,0} \p_\psi \ln q = \frac{r}{q} \frac{\p q}{\p r}
\end{equation}
which can be substituted into Eq. \eqref{eq:relationship-shear-fp} to relate the shear to $f'R_0$ as
\begin{equation}
    s = \frac{\gamma \epsilon^2}{q} f'R_0 - \frac{2}{\gamma} C_1 - \frac{2 \epsilon}{\gamma} C_2  - \frac{\alpha}{2\gamma \epsilon} C_3 + \frac{q}{\gamma^2} C_4 f'R_0,
\end{equation}
where we have defined the geometric constants $C_1$ to $C_4$ as
\begin{subequations}
\begin{eqnarray}
C_1 &=& \frac{1}{2 \upi} \oint \frac{\hat{l}_\theta}{\hat{R}_c \hat{R}_s^3 \hat{B}_{p,s}^2}  \mathrm{d}\theta, \\
C_2 &=& \frac{1}{2 \upi} \oint \frac{\hat{l}_\theta \sin u}{\hat{R}_s^4 \hat{B}_{p,s}^2}  \mathrm{d}\theta,  \\
C_3 &=& \frac{1}{2 \upi} \oint \frac{\hat{l}_\theta}{\hat{R}_s^2 \hat{B}_{p,s}^3}  \mathrm{d}\theta,  \\
C_4 &=& \frac{1}{2 \upi} \oint \frac{\hat{l}_\theta}{ \hat{R}_s^4 \hat{B}_{p,s}^3}  \mathrm{d}\theta. 
\end{eqnarray}
\end{subequations}
The radial derivatives of the magnetic field become
\begin{subequations}
\begin{eqnarray}
r \p_\rho b_p &=& \left( \frac{1}{\hat{R}_c} - \frac{\alpha}{2 \epsilon \hat{B}_{p,s}} \left[ \frac{1}{\hat{R}_s} - \hat{R}_s \right] - \frac{\sigma}{\hat{R}_s \hat{B}_{p,s}} \right), \\
r \p_\rho b_\phi &=& \epsilon \left( \frac{\gamma^2\epsilon}{q^2} \left[ \sigma + \frac{\alpha}{2 \epsilon} \right] \hat{R}_s \hat{B}_{p,s} - \frac{\sin u}{\hat{R}_s}  \right). 
\end{eqnarray}
\end{subequations}
These expressions are the same as \citet[Eq.~(14)]{Roach1995TrappedTokamaks}, where the differences in sign arise because the sign convention for $\hat{R}_c$ is different here and $\rho$ has the opposite sign. Finally, we express the total magnetic field length as
\begin{equation}
    L = \frac{q\xi}{\gamma} R_0 , \quad \xi \equiv \oint \frac{\hat{l}_\theta \hat{B}_s}{\hat{B}_{p,s}} \mathrm{d} \theta.
\end{equation}

We now turn our attention to the precession frequency, which we calculate from \eqref{eq:precession-freq}. To simplify the calculation slightly, we note that the operator $\Delta \psi_t \partial_{\psi_t} \approx \Delta \psi \partial_\psi  $ to leading order around smallness of the radial coordinate $\rho$, as we can approximate $\Delta \psi_t \approx \Delta \psi \partial_\psi \psi_t $. Using this identity, we find the same expression as in \citet{Roach1995TrappedTokamaks},
\begin{equation}
\begin{aligned}
    \hat{\omega}_\alpha(\lambda) = - \frac{\Delta \psi}{R_0^2 B_{p,0}} \Bigg\langle \frac{1}{\epsilon} \left( 2  \left[1 - \lambda \hat{B} \right] \left[r\p_\rho b - r\p_\rho b_p - \frac{1}{\hat{R}_c}\right] - \lambda \hat{B} r \p_\rho b \right) \\
    \Bigg/ \hat{B}_{p,s} \hat{R}_s \Bigg\rangle_\lambda,
    \label{eq:precession-freq-general}
\end{aligned}
\end{equation}
where we define the bounce averaging operator in angular brackets as
\begin{equation}
    \langle \dots \rangle_\lambda = \frac{\int \mathrm{d} \theta ~ \dots  \hat{l}_\theta \frac{\hat{B}_s}{\hat{B}_{p,s}} \Big/ \sqrt{1 - \lambda \hat{B}} }{ \int \mathrm{d} \theta ~ \hat{l}_\theta \frac{\hat{B}_s}{\hat{B}_{p,s}} \Big/ \sqrt{1 - \lambda \hat{B}} }.
\end{equation}
We rewrite the precession frequency as
\begin{equation}
    \hat{\omega}_\alpha \equiv - \frac{\Delta \psi}{R_0^2 B_{p,0}} \hat{\omega}_\lambda.
\end{equation}
Next, we investigate the Jacobian $\hat{g}^{1/2}$, which is the normalised bounce time, and find that it is equal to
\begin{equation}
    \hat{g}^{1/2} = \frac{\int \mathrm{d} \theta ~ \hat{l}_\theta \frac{\hat{B}_s}{\hat{B}_{p,s}} \Big/ \sqrt{1 - \lambda \hat{B}}}{\xi}
    \label{eq: dimless bounce-times}
\end{equation}
We rescale it with a factor $\epsilon^{1/2}$ to acount for the fact that in smallness of $\epsilon$ the integrand of the bounce time goes as $1/\sqrt{\epsilon}$. Therefore, we define
\begin{equation}
    \hat{g}_\epsilon^{1/2} \equiv \hat{g}^{1/2} \sqrt{\epsilon}.
    \label{eq: dimless bounce-times epsilon}
\end{equation}
The \AE{} now becomes
\begin{equation}
    A = \frac{3}{16} \frac{\Delta \psi_t \Delta \alpha_C L}{B_0} n_0 T_0  \sqrt{\epsilon} \left( \frac{\Delta \psi}{R_0^2 B_{p,0}} \right)^2 \left( \frac{1}{\epsilon} \int_{\{ \lambda \}} \mathrm{d}\lambda ~ \sum_{\text{wells}(\lambda)} I_z(c_0,c_1) \hat{\omega}_\lambda^2 \hat{g}_\epsilon^{1/2} \right),
\end{equation}
where the prefactor $1/\epsilon$ to the integral deliberately not cancelled against the $\sqrt{\epsilon}$, so that the integral in brackets is to lowest order independent of $\epsilon$, as the integration range scales as $\epsilon$. With the above expression, we go on to define a dimensionless \AE{}. We take steps in accordance with \citep{Mackenbach2022AvailableTransport}, and calculate the fraction of the total thermal energy that is available. The thermal energy of a plasma in a flux tube can be calculated by expanding around $\psi_t = \psi_{t,0}$ and $\alpha_C = \alpha_{C,0}$ and retaining only the constant terms, resulting in
\begin{equation}
    E_t = \int \frac{3}{2} \frac{n T}{B} \mathrm{d} \psi_t \mathrm{d} \alpha_C \mathrm{d} \ell = \frac{3}{2} n_0 T_0 \frac{ \Delta \psi_t \Delta \alpha_C L }{B_0} \frac{1}{\xi} \oint \hat{l}_\theta \hat{B}_{p,s}^{-1} \mathrm{d} \theta.
\end{equation}
We then define the available energy as a fraction of the thermal energy as
\begin{equation}
    \widehat{A} = \frac{A}{E_t}.
    \label{eq:AE-tok-general}
\end{equation}
Simplifying the expression using $\Delta \psi = \Delta r \partial_r \psi$, one finds that
\begin{equation}
    \widehat{A} = \frac{1}{8} \left( \frac{\Delta r}{R_0} \right)^2 \frac{\xi \sqrt{\epsilon}}{\oint \hat{l}_\theta \hat{B}_{p,s}^{-1} \mathrm{d} \theta } \cdot \frac{1}{\epsilon} \int_{\{ \lambda \}} \mathrm{d}\lambda ~ \sum_{\text{wells}(\lambda)} I_z(c_0,c_1) \hat{\omega}_\lambda^2 \hat{g}_\epsilon^{1/2}.
    \label{eq:ae-final}
\end{equation}
$\Delta r$ measures the length-scale over which energy is available, i.e. a typical length-scale over which gradients can be flattened. We take this to be proportional to the correlation length, typically found to be the gyroradius. Therefore, let us set 
\begin{equation}
    \Delta r = C_r \rho_\text{g},
    \label{eq:AE-lengthscale}
\end{equation}
where $\rho_\text{g}$ is the gyroradius, and $C_r$ some function of order $\mathcal{O}(\rho_\text{g}^0)$. This function is not known \emph{a priori}, and may vary. For example, if there are large radial streamers present in the system $C_r$ may be significantly increased. The dimensionless \AE{} now becomes
\begin{equation}
    \widehat{A} = \frac{1}{8} \left( \frac{\rho_\text{g}}{R_0} \right)^2 \frac{ C_r^2 \xi \sqrt{\epsilon}}{\oint \hat{l}_\theta \hat{B}_{p,s}^{-1} \mathrm{d} \theta } \cdot \frac{1}{\epsilon} \int_{\{ \lambda \}} \mathrm{d}\lambda ~ \sum_{\text{wells}(\lambda)} I_z(c_0,c_1) \hat{\omega}_\lambda^2 \hat{g}_\epsilon^{1/2}.
    \label{eq:ae-final-with-Cr}
\end{equation}
This expression has various scalings which are of interest. First, we see that reducing the aspect ratio for fixed $\rho_\mathrm{g}/R_0$ is beneficial since it leads to fewer trapped particles. Note that, in the limit of a large aspect ratio, the trapping fraction scales as $\sqrt{\epsilon}$, which is the same dependency found here. A reduction in the expansion parameter $\rho_\text{g}/R_0$ (at fixed $\epsilon$) is also found to help decrease \AE{}.
\par
As a final step, we introduce the dimensionless density gradient
\begin{equation}
    R_0^2 B_{p,0} \p_\psi \ln n = \frac{R_0}{n} \frac{\partial n}{\partial r} \equiv - \hat{\omega}_n,
    \label{eq: gradient strength}
\end{equation}
with which $c_0$ and $c_1$ reduce to an especially simple form
\begin{equation}
    c_0 = \frac{\hat{\omega}_n}{\hat{\omega}_\lambda} \left( 1 - \frac{3}{2}\eta \right), \quad c_1 = 1 - \frac{\hat{\omega}_n}{\hat{\omega}_\lambda} \eta.
    \label{eq:c1c2-definition}
\end{equation}
Importantly let us make note of the fact that the radial coordinate $r$ may have different conventions. In previous investigations \citep{Mackenbach2022AvailableTransport,Mackenbach2023AvailableTransport} the radial coordinated was defined via the square root of the toroidal flux
\begin{equation}
    r_{\rm eff} \propto \sqrt{\psi_t},
\end{equation} 
with $\psi_t$ being the toroidal flux passing through flux surface in question. A different choice of the radial coordinate $r$ will influence various quantities on which \AE{} depends, such as Eqs. \eqref{eq:AE-lengthscale} and \eqref{eq: gradient strength}. More specifically, the length scale $\Delta r$ expressed in terms of $r_{\rm eff}$ is
\begin{equation}
    \Delta r = \Delta r_{\rm eff} \frac{\partial r}{\partial r_{\rm eff}}.
\end{equation}
In the aforementioned investigations $\Delta r_{\rm eff}$ was chosen as $\Delta r_{\rm eff} = \rho $, resulting in a good correlation with turbulent energy fluxes. Therefore, we choose $C_r$ such that $\Delta r_{\rm eff} = \rho$, which means that
\begin{equation}
    C_r = \frac{\partial r}{\partial r_{\rm eff}},
\end{equation}
and we shall use this choice of $C_r$ from here on.

\subsection{Miller geometry}
\begin{figure}
    \centering
    \includegraphics[width=\textwidth]{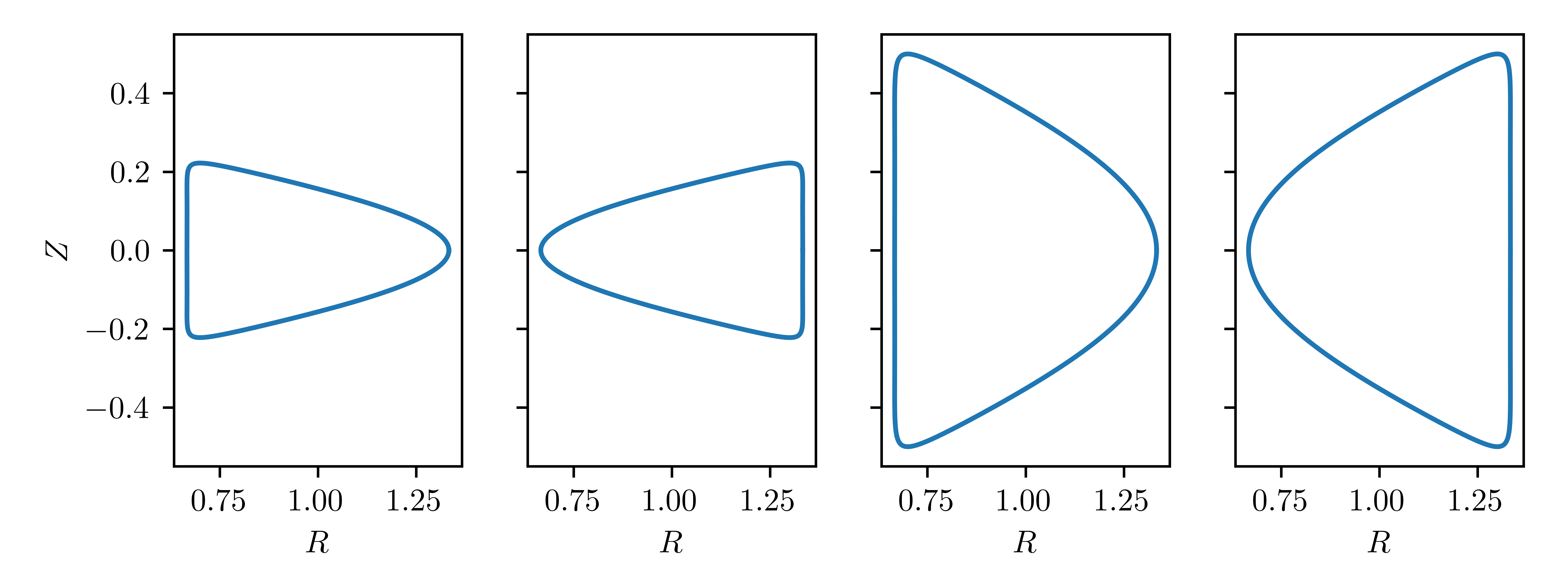}
    \caption{Cross sections in the $(R,Z)$-plane of tokamaks parameterised via Eq. \eqref{eq: Miller param}. The parameters $(\kappa,\delta)$ vary from the plot on the left to the plot on the right, as $(2/3,0.9)$, $(2/3,-0.9)$, $(3/2,0.9)$, and $(3/2,-0.9)$. All plots have $R_0=1$ and $\epsilon=1/3$.}
    \label{fig:Miller reference shapes}
\end{figure}

Finally, we choose our equilibrium to be of the type discussed in \citet{Miller1998NoncircularModel}. The key step is to parameterise the flux surface as a standard D-shaped tokamak in terms of the poloidal angle $\theta$,
\begin{subequations}
\begin{eqnarray}
R_s(\theta) &=& R_0 + R_0 \epsilon \cos(\theta +  \arcsin [\delta] \sin \theta), \\
Z_s(\theta) &=& R_0 \kappa \epsilon \sin \theta.
\end{eqnarray}
\label{eq: Miller param}
\end{subequations}
Here, $R_0(r)$ is the centre of the flux surface, $\kappa(r)$ is the elongation, and $\delta(r)$ is the triangularity. An important feature of this parameterisation is that it is up-down symmetric, which can be seen by invariance under $(Z_s,\theta)\mapsto-(Z_s,\theta)$. The poloidal magnetic field can then be calculated by \eqref{eq:poloidal-field}, and the equilibrium is fully specified by the following set of 9 parameters; $[\epsilon,\kappa,\delta, s_\kappa, s_\delta, \p_r R_0, q, s, \alpha]$, where $s_\kappa =r \p_r \ln \kappa$ and $s_\delta = r \p_r \arcsin(\delta)$. Henceforth we shall refer to this set of numbers which determines the local geometry as a ``Miller vector'',
\begin{equation}
    \boldsymbol{M} = [\epsilon,\kappa,\delta, s_\kappa, s_\delta, \p_r R_0, q, s, \alpha].
    \label{eq:miller-vector}
\end{equation}
Cross-sections are plotted in Fig. \ref{fig:Miller reference shapes}, to serve as a reference for the various shapes mentioned in the following sections.
We furthermore recognise that it is possible to calculate the toroidal flux enclosed by a poloidal cross-section, and one may retrieve analytical expressions by expanding it around the smallness of $\epsilon$,
\begin{equation}
\begin{aligned}
    \psi_t &= B_0 \pi r^2 \frac{2}{\pi} \int_{\pi}^{0} \hat{Z}_s \frac{\partial_\theta R_s}{r\hat{R}_s} \mathrm{d} \theta \\
    &\approx B_0 \pi r^2 \kappa  \left( \frac{2 J_1(\arcsin \delta)}{\arcsin \delta} +  \frac{J_2(2\arcsin \delta) }{2\arcsin \delta} \epsilon + \mathcal{O}(\epsilon^2) \right),
\end{aligned}
\end{equation}
with $J_n(x)$ being the $n^{\rm th}$ Bessel function of the first kind. In terms of an effective $r$, we then have
\begin{equation}
\begin{aligned}
    r_{\rm eff} &= r \sqrt{ \frac{2}{\pi} \int_{\pi}^{0} \hat{Z}_s \frac{\partial_\theta R_s}{r\hat{R}_s} \mathrm{d} \theta }  \\
    &\approx r \sqrt{\kappa} \sqrt{\frac{2 J_1(\arcsin \delta)}{\arcsin \delta}} \left( 1 + \frac{1}{8} \frac{J_2 (2 \arcsin \delta)}{J_1 (\arcsin \delta)} \epsilon + \mathcal{O}(\epsilon^2) \right),
\end{aligned}
\label{eq: C_r limit}
\end{equation}
and we find that the factor $C_r$ becomes
\begin{equation}
    C_r= \left(\sqrt{\frac{2}{\pi} \int_{\pi}^{0} \hat{Z}_s \frac{\partial_\theta R_s}{r\hat{R}_s} \mathrm{d} \theta} \right)^{-1}.
\label{eq: C_r choice}
\end{equation}
For shaped equilibria (i.e. $\kappa \neq 1$, $\delta \neq 0$, or $\epsilon \rightarrow 1$), $C_r$ will differ from unity and one should keep this important caveat in mind.
\subsection{An analytical limit: large aspect ratio $s$-$\alpha$ tokamak}
We proceed to investigate a limiting case of Miller geometries; namely that of a large aspect ratio tokamak with circular flux surfaces and a steep local pressure gradient, which we shall henceforth refer to as the $s$-$\alpha$ limit, and this calculation is equivalent to analyses given in \citet{Connor1983EffectTokamak,Roach1995TrappedTokamaks}. This will serve as a computationally efficient model in such geometries, and will furthermore be used as a benchmark for the more general calculation of the \AE{}. The algebraic details of this derivation are given in Appendix \ref{sec:details-derivation-of-s-alpha-integrals}, and here we highlight the central steps. It is convenient to express $\lambda$ as a trapping parameter $k^2$, where the deeply trapped particles have $k=0$ and the barely trapped particles have $k=1$. This mapping is given by $\lambda = 1 + \epsilon (1 - 2 k^2)$, so the magnetic field may be written as
\begin{equation}
    \lambda \hat{B} = 1 + \epsilon (1 - 2 k^2 - \cos \theta).
\end{equation} 
One can now perform the bounce-averaging integrals required for Eq. \eqref{eq:precession-freq-general} in the $s$-$\alpha$ limit, resulting in
\begin{equation}
    \hat{\omega}_\lambda = -\frac{\alpha}{2q^2} + 2 G_1(k) + 4s G_2(k) - \alpha G_3(k),
    \label{eq: w CHM}
\end{equation}
where we define
\begin{subequations}
\begin{eqnarray}
G_1 &=& \frac{E(k)}{K(k)} - \frac{1}{2}, \\
G_2 &=& \frac{E(k)}{K(k)} + k^2 - 1,  \\
G_3 &=& \frac{2}{3} \left[ \frac{E(k)}{K(k)} (2k^2 - 1) + 1 - k^2 \right],
\end{eqnarray}
\end{subequations}
where $K(k)$ and $E(k)$ are complete elliptic integrals of the first and second kind, respectively. The normalised bounce time, as given in \eqref{eq: dimless bounce-times epsilon} is equal to
\begin{equation}
    \hat{g}_\epsilon^{1/2} = \frac{\sqrt{2}}{\pi} K(k).
\end{equation}
Finally, from Eq. \eqref{eq: C_r limit} we see that $C_r = 1$ in this limit. The \AE{} now becomes a straightforward integral of known functions over $k$
\begin{equation}
    \widehat{A} = \frac{1}{2\upi\sqrt{2}}  \left( \frac{\rho_\text{g}}{R_0} \right)^2 \sqrt{\epsilon} \int_0^1 \mathrm{d} k^2 ~ I_z(c_0(k),c_1(k)) \hat{\omega}_\lambda(k)^2 K(k),
    \label{eq:s-alpha-exact}
\end{equation}
which can efficiently be computed numerically.

\section{Numerical results}
Two codes have been constructed: one that computes the integral of \eqref{eq:s-alpha-exact} using standard integration routines, and a numerical routine that computes both the precession frequencies and the \AE{} as given in \eqref{eq:AE-tok-general}, both of which are computationally cheap (fractions of a CPU second per evaluation). First, we shall verify the relationship between \AE{} and turbulent transport. Next, we shall investigate the results obtained for the $s$-$\alpha$ circular tokamak, after which we shall investigate how \AE{} varies in Miller geometries as a function of various parameters. The code used to generate these results is freely available on GitHub\footnote{Install the code via \url{https://github.com/RalfMackenbach/AE-Miller}}. The bounce-integrals required in Eq. \eqref{eq:AE-tok-general} are evaluated using numerical methods detailed in \citet{mackenbach2023drift}. Finally, we take the prefactor $\rho_\text{g}/R_0$ to be unity in all plots presented below, so when converting to a real device, one should multiply the \AE{} by a factor $(\rho_\text{g}/R_0)^2$.
\subsection{Comparison with \textsc{tglf}}
Our first course of action is comparing \AE{} with turbulent energy-flux calculations in tokamak geometries, to verify its relation to turbulent transport in such geometries. At the moment, nonlinear gyrokinetic simulations are computationally too expensive for detailed parameters scans, and therefore we instead employ the quasi-linear \textsc{tglf} (trapped gyro-Landau fluid) code \citep{staebler2007theory,staebler2010electron}.
Some key differences between the two models are highlighted before any comparison is made.
\textsc{tglf} computes the linear eigenmodes of a variety of instabilities, ion and electron temperature gradient (ITG, ETG) modes, electromagnetic kinetic ballooning (KB) modes, as well as trapped-ion and trapped-electron modes (TIM, TEM), and then applies a quasilinear saturation rule to accurately fit the fluxes from nonlinear gyrokinetic simulations. For quasi-neutrality purposes, \textsc{tglf} requires the inclusion of at least one ion species. These are fundamental differences to the formulation of the \AE{} described in this work, which only accounts for the \AE{} of trapped electrons. Therefore, when setting up \textsc{tglf}, care was taken to ensure the modelled turbulent energy-fluxes were as much as possible due to instabilities dominated by trapped electrons, using settings analogous to those used in recent gyrokinetic simulations in a similar regime \citep{Proll2022mitigation}. Given the lack of collisions in this regime, the expected dominant instabilities should be of the collisionless trapped-electron mode (CTEM) variety. However, some other instabilities can also arise from interactions with the ion population. Thus, to ensure that the dominant instabilities in the \textsc{tglf} simulations were as relevant as possible for our comparison, only contributions from modes propagating in the electron-diamagnetic direction were included, which excludes e.g. the ubiquitious mode \citep{coppi1977theory}, which propagates in the ion direction. Furthermore, we find that, for the scenarios considered in this work, adding an equally large electron temperature gradient to the density gradient, i.e. taking $\eta=1$, significantly decreased the amount of non-TEM modes dominant in \textsc{tglf} simulations, and as such we set $\eta$ to unity for the comparison. The recent SAT2 \citep{staebler2021verification} quasilinear saturation rule for \textsc{tglf} was used, as it includes the impact of plasma shaping on the quasilinear saturation \citep{staebler2020geometry}.
Although \textsc{tglf} also uses a Miller parameterisation of the local equilibrium, we note that it does not use the same normalisation as \citet{Roach1995TrappedTokamaks} followed in this work, and care has been taken to convert between the two. We finally stress that the current model for $C_r$ is a fairly simple model, and that prediction can be refined using a more sophisticated model. This can, for example, be done by using some fitting function for $C_r$, where one finds the best-fit parameters which minimise the error between the energy flux and the prediction of \AE{}.
\begin{figure}
    \centering
    \includegraphics[width=\textwidth]{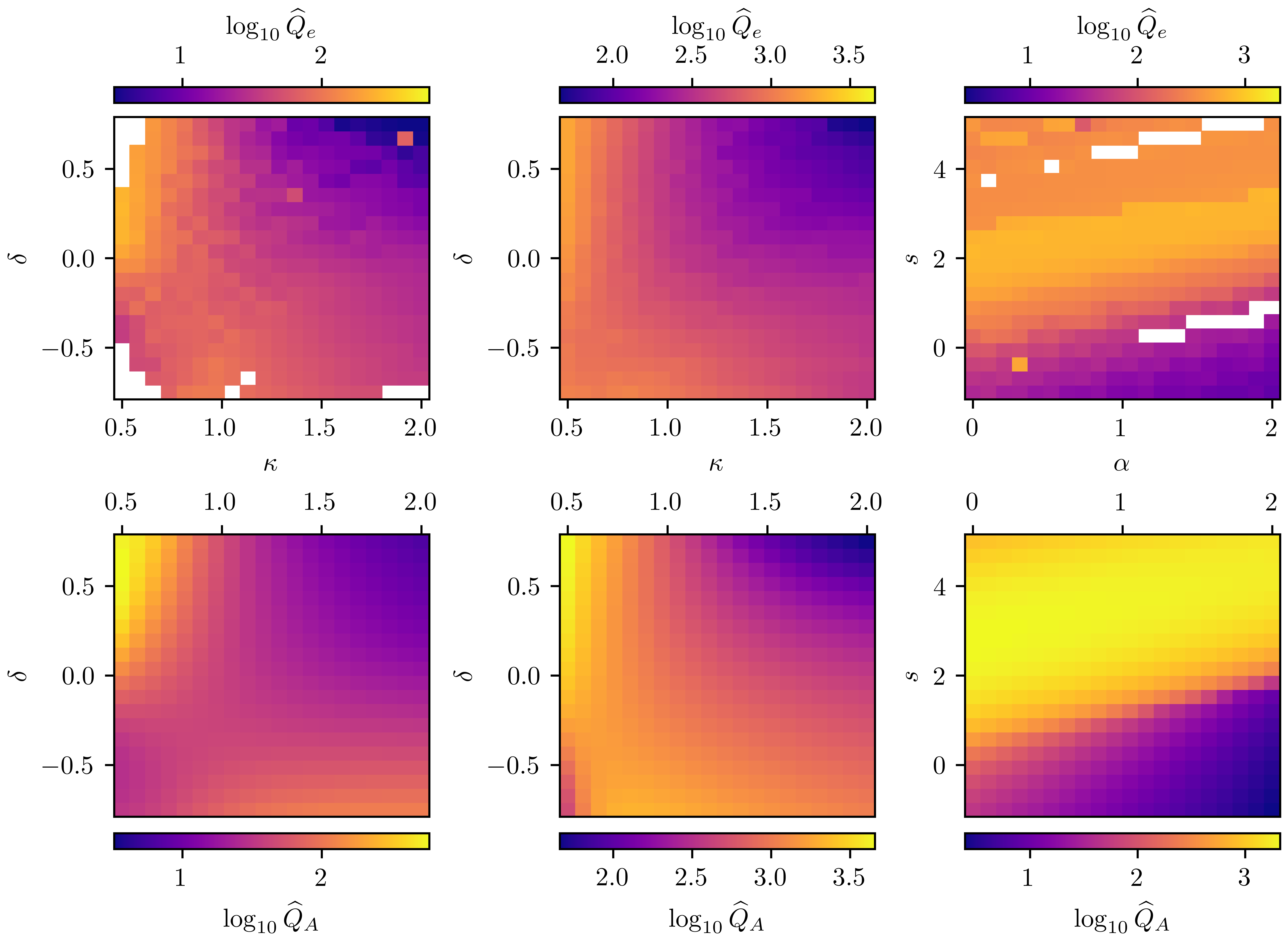}
    \caption{Comparison between codes, showing the correspondence between the estimate of the energy flux from \AE{} and \textsc{tglf}. The top row displays the energy flux from \textsc{tglf}, and the bottom row displays the corresponding estimate from \AE{}. One can see agreement in trends, though some details differ. The left column has a Miller vector $[\epsilon,\kappa,\delta, s_\kappa, s_\delta, \p_r R_0, q, s, \alpha]$ of $ [1/3,\kappa,\delta, 0, 0, 0, 2, 0, 0]$, whereas the middle column has $ [1/3,\kappa,\delta, 0, 0, 0, 2, 1, 0]$, and the right-most column has as Miller vector of $ [1/3,1,1/2, 0, 0, 0, 2, s, \alpha]$. The first column has $\hat{\omega}_n=3$, the second and third column have $\hat{\omega}_n=6$, and $\eta=1$ in all plots. The white masked-out regions have a dominant instability which is not in the electron direction, and as such we filter them out. Finally, note that the colour bars are the same scale in each column.}
    \label{fig:tglf-comparison}
\end{figure}
\par 
For the comparison we use the gyro-Bohm normalised energy fluxes computed by \textsc{tglf},
\begin{equation}
    \widehat{Q}_e = \frac{Q_e}{Q_\mathrm{GB}},
\end{equation}
where $Q_e$ is the electron energy flux from \textsc{tglf}, and $Q_\mathrm{GB}$ is the gyro-Bohm energy flux. This is compared to the estimate of the gyro-Bohm normalised energy flux from \AE{} \cite{Mackenbach2022AvailableTransport,Mackenbach2023AvailableTransport}, which is
\begin{equation}
    \widehat{Q}_{A} \equiv C_Q \widehat{A}^{\: 3/2},
\end{equation} 
where the constant of proportionality is taken from the fit presented there we was found $C_Q \approx  1.0 \cdot 10^3$. With such a power law, a linear correlation between $\widehat{Q}_A$ and $\widehat{Q}_e$ from nonlinear gyrokinetic simulations was found for pure density gradient-driven TEMs, which is different from the current comparison in which both the electron temperature and the density gradient drive the TEM ($\eta=1$). The data points in the comparison are chosen in order to verify that \textsc{tglf} reproduces some trends that will be discussed in following sections.
\begin{figure}
    \centering
    \includegraphics{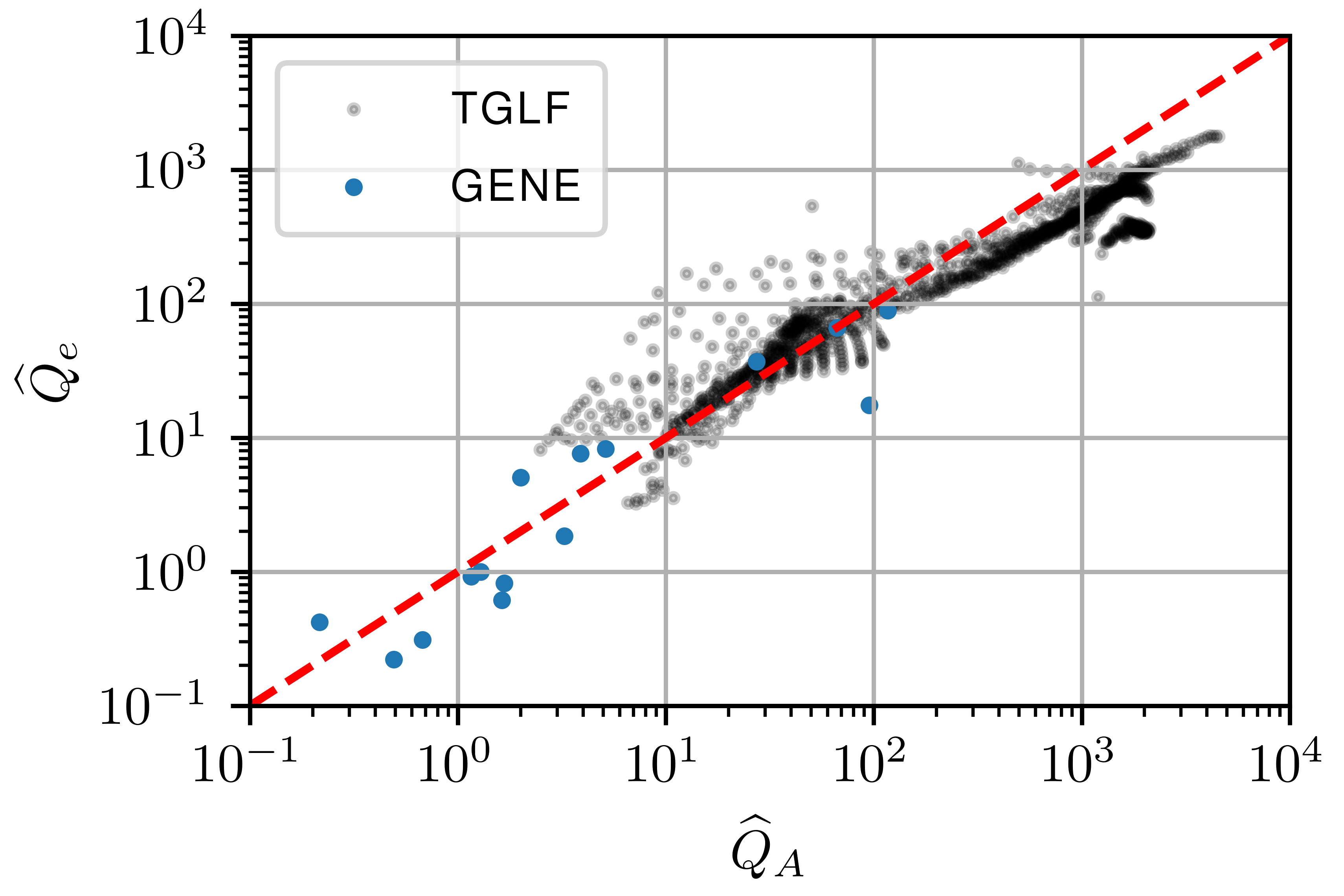}
    \caption{A scatter plot showing the relation between the two estimates of the non-linear energy flux. The red dashed line is shows the expected linear relationship, $x=y$. The plot consists of $N=1171$ points. The gray points have some transparency, and as such darker regions arise due to a high density of points. We have furthermore added simulation data from the gyrokinetic code \textsc{gene} \citep{jenko2001critical}, also presented in \citet{Mackenbach2022AvailableTransport}, as blue markers.}
    \label{fig:scatter-tglf-ae}
\end{figure}
\par
A comparison in the $(\kappa,\delta)$ and $(s,\alpha)$ planes is displayed in Fig. \ref{fig:tglf-comparison}. One can see that there is good correspondence in trends: decreasing the magnetic shear and/or increasing the pressure gradient helps in reducing the energy flux, as does increasing the elongation. However, there are also differences visible between the two models for the energy flux, which are evident in the $(s,\alpha)$-plot. A clear discrepancy can be seen at high shear values $(s \approx 3)$, where the \textsc{tglf} energy flux drops and the \AE{} estimate does not, and the \AE{} furthermore overestimates the energy flux at high shear. In the $(\kappa,\delta)$-plots the trends are well captured by \AE{}, with some differences. To further investigate the relationship between the two estimates of the energy flux, all the simulation data shown in Fig. \ref{fig:tglf-comparison} have been combined in a scatter plot shown in Fig. \ref{fig:scatter-tglf-ae}. In order to check consistency with previous findings, we have furthermore included the data points of \citet{Mackenbach2022AvailableTransport}, which are nonlinear simulations in general geometries. Here, we see that there is a linear relationship for most of the data (we have added a red line with the expected linear relationship), although there exist data points that deviate more significantly from the linear relationship. There are various reasons why such a discrepancy may occur:
\begin{itemize}
    \item There may be other instabilities present (though not necessarily dominant) that are not captured by the \AE{} of trapped electrons, such as the ubiquitous mode, or the universal instability \citep{landreman2015universal,helander2015universal,romanelli1989ion,costello2023universal}. More generally, if there are instabilities present that do not derive their energy from trapped electrons, the current \AE{}-model is no longer expected to be an accurate measure.
    \item The \AE{} length-scale $C_r$ may vary more significantly for certain choices of equilibrium parameters and the current choice given in Eq. \eqref{eq: C_r choice} may not be accurate.
    \item Recall that \AE{} can be interpreted as an {\it upper bound} on the amount of energy that can be released. If the portion of the \AE{} that resides in stabilising modes deviates markedly (see, e.g. \citet{lang2008nonlinear,hatch2011role,pueschel2016stellarator,duff2022effect}), one can reasonably expect that the data deviate more from the found relationship. 
    \item The \textsc{tglf}'s quasilinear saturated fluxes in both the $(\kappa,\delta)$ and $(s,\alpha)$ planes show occasional extreme outliers for small changes in input. \textsc{tglf} has been extensively verified against a wide variety of nonlinear gyrokinetic simulations (although further validation for negative triangularity is currently being pursued), but the regime explored in this work is not the typical input space and could require separate verification.
    \item Although not present in the current set of simulation data, the \AE{} of trapped electrons will certainly cease to be an accurate model in situations where the trapped electrons play no role, such as in the case of a pure ion temperature gradient, and no gradients in of electron temperature or density.
\end{itemize}
\par
We stress that the scatter does mean that predictions may be faulty if one lies within the scatter of the fit. However, seeing that general trends are well captured by \AE{}, it may serve as a useful estimate for transport and trends at low computational cost (\AE{} calculations are roughly a factor $50$ faster than the presented \textsc{tglf} calculations).
\subsection{$s$-$\alpha$ geometry} \label{sec:s-alpha-results}
\begin{figure}
    \centering
    \includegraphics[width=\textwidth]{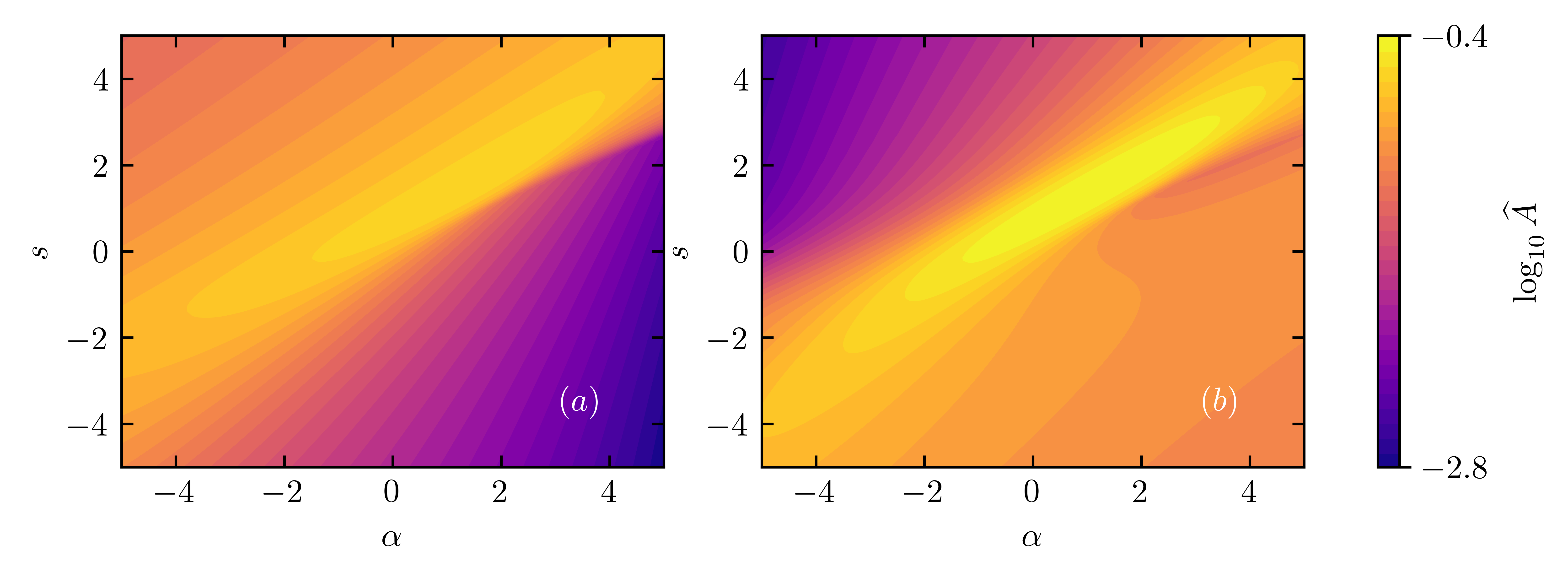}
    \caption{The \AE{} of a large aspect ratio circular tokamak, as a function of magnetic shear $s$ and pressure gradient $\alpha$. The plots have been generated using $q=2$. Plot $(a)$ has $\hat{\omega}_n=3$ and $\eta=0$, whereas plot $(b)$ has a pure electron temperature gradient, i.e. $\hat{\omega}_n=0$ and $\hat{\omega}_n \cdot \eta = 3$.}
    \label{fig:s-alpha-circ}
\end{figure}
We now shift our attention to the behaviour of \AE{} on the various free parameters found in tokamak equilibria. Recalling that we have derived two \AE{} expressions, one for any Miller geometry and one for the large-aspect ratio limit, let us start by investigating the latter. A plot of the \AE{} calculated from Eq. \eqref{eq:s-alpha-exact} is given in Fig. \ref{fig:s-alpha-circ} as a function of magnetic shear and pressure gradient. We note that the ranges for $s$ and $\alpha$ are not meant to represent realistically attainable values here, instead, we are more interested in the general structure of the \AE{} over the domain. There are several interesting features visible in the figure. Even in this simplest model, the available energy exhibits rich structure over the $s$-$\alpha$ plane. More precisely, \AE{} is large when $s$ and $\alpha$ are comparable, $s\sim\alpha$, and is otherwise much smaller, particularly when the absolute value of one of these quantities is large. These findings are consistent with previous investigations \citep{rosenbluth1971finite,Dagazian1982TheBallooning,Connor1983EffectTokamak,Kessel1994ImprovedShear,Strait1997StabilityTokamak,Rettig1997MicroturbulenceDischarges,Kinsey2006TheSimulations}. It is also interesting to note that the precise reduction in \AE{} depends on the drive: for a pure electron temperature gradient, significant positive shear is more helpful in reducing \AE{}, while \AE{} driven by a pure density gradient benefits more from negative shear. \par 
Since Eq. \eqref{eq:s-alpha-exact} can be integrated numerically to high precision, it serves as a useful benchmark for the more general \AE{} of \eqref{eq:AE-tok-general}. Accordingly, we have compared the \AE{} in the large-aspect-ratio limit with circular flux surfaces using a code that solves Eq. \eqref{eq:AE-tok-general}. This comparison is shown in Appendix \ref{sec:appendix-salpha-miller-comparison}, and we find that the codes agree.
\subsection{Miller geometry}
\begin{figure}
    \centering
    \includegraphics[width=\textwidth]{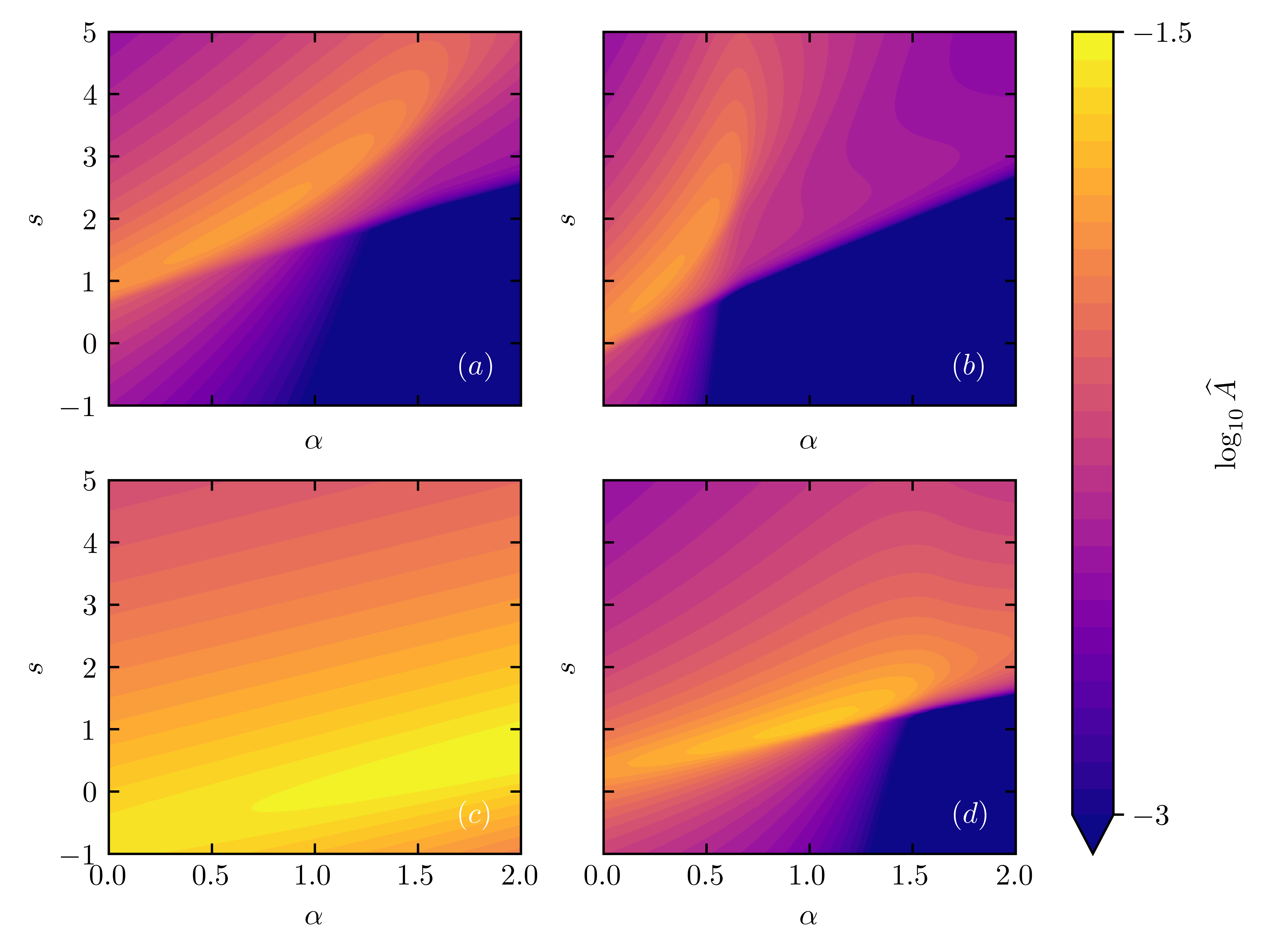}
    \caption{Dependence of \AE{} on magnetic shear $s$ and pressure gradient $\alpha$. In subplot $(a)$, the Miller vector $[\epsilon,\kappa,\delta, s_\kappa, s_\delta, \p_r R_0, q, s, \alpha]$ is set to $\left[ 1/3, 3/2, 1/2, 0, 0, 0, 2, s,\alpha \right]$, and other plots have the same vector with one parameter changed. In the subplot $(b)$ the safety factor $q$ is reduced from $2$ to $1$, for $(c)$ the elongation $\kappa$ decreases from $3/2$ to $1/2$, and in $(d)$ the sign of triangularity $\delta$ is changed from $1/2$ to $-1/2$. All plots have $\hat{\omega}_n=1$ and $\eta = 0$.}
    \label{fig:scan-salpha}
\end{figure}

We now leave the realm of the $s$-$\alpha$ limit and venture into shaped, finite-aspect-ratio equilibria. Our first step is to investigate the dependence on magnetic shear and pressure gradient for a range of different Miller vectors, and the results are shown in Fig. \ref{fig:scan-salpha}. 
Here we see similar trends as in section \ref{sec:s-alpha-results}: negative shear and large $\alpha$ tend to be especially stabilising for a pure density gradient. However, it is also clear that the magnitude and precise contours depend \emph{strongly} on the chosen Miller vector, as defined in Eq. \eqref{eq:miller-vector}. For example, it can be seen that lowering the safety factor is stabilising, since \AE{} is reduced over a large region of the $s$-$\alpha$ plane as one compares subfigure $(a)$ to $(b)$. In subfigure $(c)$ the elongation has been reduced  produce a ``comet''-type configuration ($\kappa < 1$, i.e. a horizontally elongated tokamak, see Fig. \ref{fig:Miller reference shapes}), which can increases the magnitude of the \AE{}, and the stabilising effects of $s$ and $\alpha$ become less pronounced. Finally, in subfigure $(d)$ the sign of the triangularity has been reversed to become negative. Although the shape of the contours remains largely unchanged, the peak in \AE{} is changed to higher $\alpha$ and lower $s$, indicating that negative triangularity can be particularly beneficial in high-shear discharges with a modest value for $\alpha$. In a more general sense, when changing any of the parameters significantly, one should expect that the precise shape and magnitude of the contours will change. \par 
\begin{figure}
    \centering
    \includegraphics[width=\textwidth]{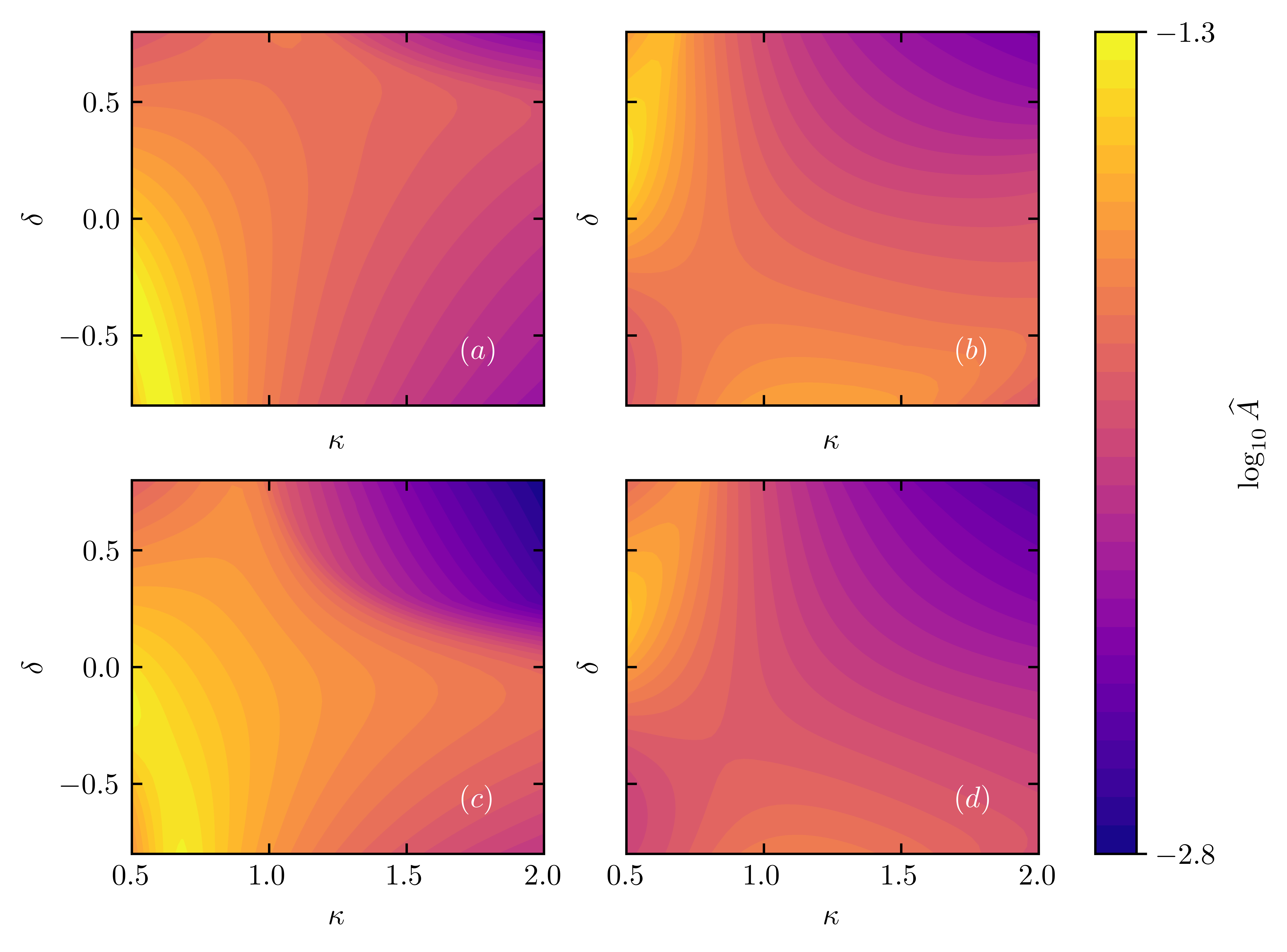}
    \caption{The effect of the geometry on \AE{}. In plot $(a)$ the Miller vector $[\epsilon,\kappa,\delta, s_\kappa, s_\delta, \p_r R_0, q, s, \alpha]$ is set to $\left[ 1/3,\kappa, \delta, 0, 0, 0, 2, 1,0 \right]$ with $\hat{\omega}_n=1$ and $\eta = 0$, and all other subfigures have the same Miller vector with one parameter changed. The contours shown in $(b)$ have a higher inverse aspect ratio as $\epsilon$ is increased from $1/3$ to $2/3$, in $(c)$ the pressure gradient $\alpha$ is increased from $0$ to $1/2$, and for $(d)$ we have decreased the shear from $1$ to $0$. }
    \label{fig:scan-geom}
\end{figure}
With this important caveat in mind, let us investigate the influence of geometry on the \AE{}. To do so, we display the dependence on $\kappa$ and $\delta$ for various Miller vectors in Fig. \ref{fig:scan-geom}. Several interesting general trends can be observed. First, note that increasing the elongation beyond $\kappa = 1$ generally decreases the \AE{} in all Miller vectors considered here, although the precise effect depends on the triangularity. Second, we see that it is not true \emph{in general} that positive or negative triangularity is always stabilising; it depends on the other Miller parameters. Third, we see that tokamaks with $\kappa <1$ and $\delta < 0$, often referred to as (negative) comet cross sections tokamaks \citep{Kesner1995CometTokamaks}, show a reduction in \AE{} in plots $(b)$ and $(d)$, at least for the pure density gradient considered here. This is perhaps unsurprising, since such tokamaks are close to having the maximum-$\mathcal{J}$ property as shown by \cite{Miller1989MaximumShaping}. Since \AE{} measures deviations from the maximum-$\mathcal{J}$ property, it is thus expected that these configurations perform well in terms of \AE{}. \par
Investigating the plots in detail, in plot $(a)$ one sees that negative triangularity is beneficial for $\kappa > 1$ as can be seen by the reduction in \AE{}. In the following sections, we shall see that this is a consequence of the positive magnetic shear chosen. Next, note that doubling the inverse aspect ratio, as is done when going from $(a)$ to $(b)$, has a stabilising effect. Naively, one would expect that doubling the inverse aspect ratio would increase the \AE{} by roughly a factor $\sqrt{2} \approx 1.4$, due to the factor $\sqrt{\epsilon}$ in Eq. \eqref{eq:ae-final}. However, going from plot $(a)$ to $(b)$ we see a {\it decrease} of the maximum \AE{} by some $15 \%$. This is likely due to the fact that, in a small-aspect-ratio device, magnetic field lines spend most of their time (or more precisely, arc-length) on the inboard side of the tokamak \citep{Helander2005CollisionalPlasmas}. There, $\omega_\lambda$ tends to be opposite to the drift wave and therefore these orbits do not contribute to the \AE{} for a pure density gradient. It is also interesting to note that negative triangularity no longer exhibits a reduction in \AE{} as the aspect ratio is significantly decreased, in accordance with the findings of \citet{balestri2023aspectplasmas}. Going from plot $(a)$ to $(c)$ the pressure gradient is increased from $\alpha = 0$ to $1/2$. With this introduction of pressure gradient, it can be seen that positive triangularity shows a decrease in \AE{}, where negative triangularity does not. Finally, plot $(d)$ has the magnetic shear reduced from $s=1$ to $s=0$ as compared to $(a)$, which drastically changes the picture. Most importantly, we see that the lack of this positive magnetic shear results in negative triangularity no longer being stablising. We find that the results change somewhat if one instead imposes a pure electron temperature gradient (not shown here), though the basic trends remain intact. \par
All in all, we conclude from these results that the \AE{} is very sensitive to equilibrium parameters, including quantities not investigated here such as $q$, $s_\kappa$, and $\partial_r R_0$. This sensitivity is perhaps reassuring: gyrokinetic turbulence has long been known to be strongly dependent on equilibrium parameters and even slight nudges can drastically change the picture (a sentiment perhaps best captured by the old Dutch expression \emph{wie het kleine niet eert, is het grote niet weerd}). We seem to reproduce a similar sensitivity in this \AE{}-model for trapped electrons. This sensitivity becomes especially clear when investigating the dependence of \AE{} on triangularity, which we shall discuss in the next section.
\subsection{When is negative triangularity beneficial?} \label{sec:neg-triang-ben}
\begin{figure}
    \centering
    \includegraphics[width=\textwidth]{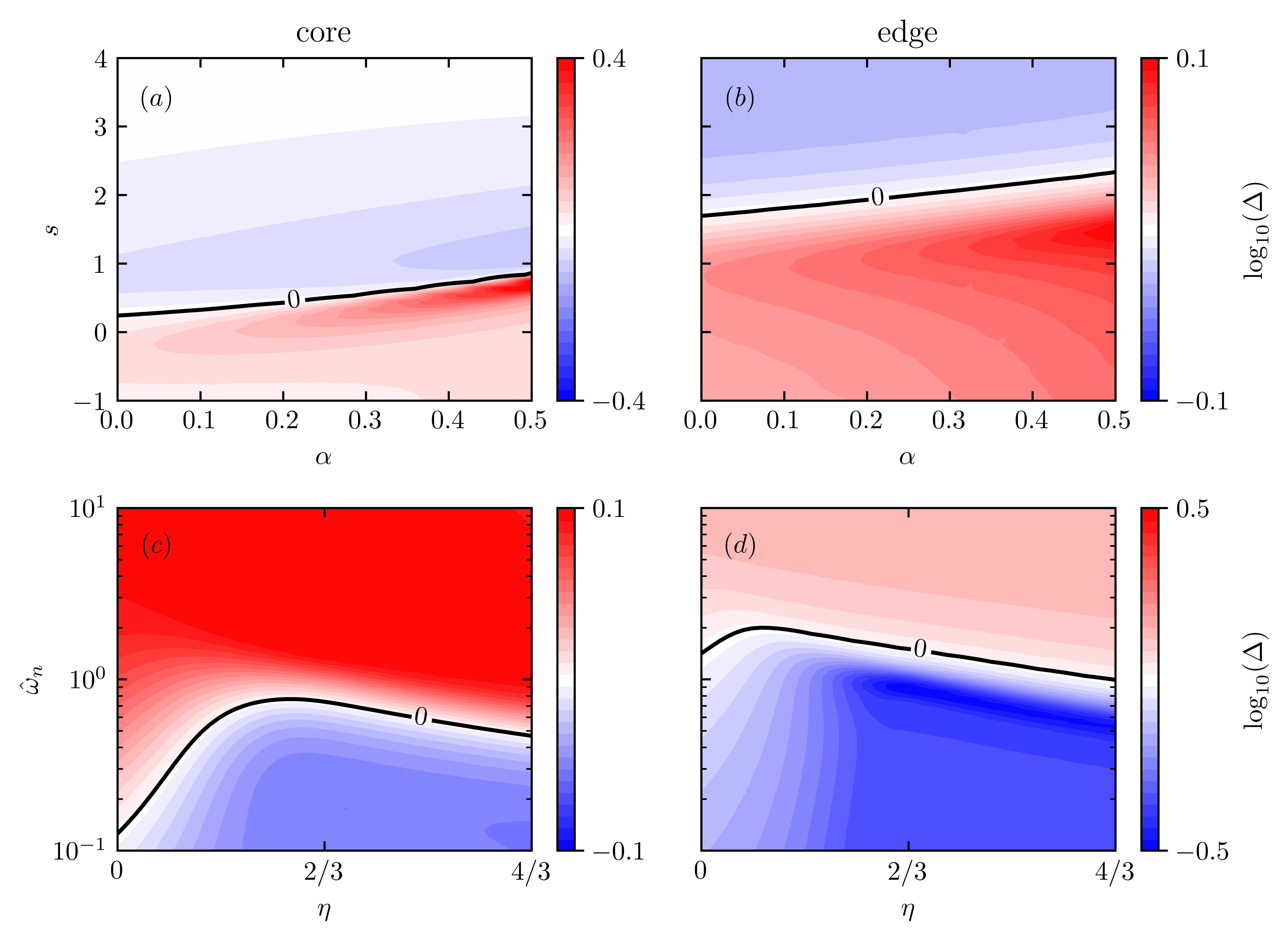}
    \caption{Various plots showcasing the dependencies of $\log(\Delta)$, where $\Delta$ is defined as $\widehat{A}(\delta=-0.1)/\widehat{A}(\delta=+0.1)$, on various equilibrium parameters. The two columns have different Miller vectors $[\epsilon,\kappa,\delta, s_\kappa, s_\delta, \p_r R_0, q, s, \alpha]$, which are meant to be representative of the edge and the core. The first column has a core-like Miller vector of $\boldsymbol{M}_\mathrm{core}=[1/100,3/2,\delta,0,0,0,1,s,\alpha]$. The second column has an edge-like Miller vector of $\boldsymbol{M}_\mathrm{edge}=[1/3,3/2,\delta,1/2,0,-1/2,3,s,\alpha]$ Finally, plot $(a)$ has $\hat{\omega}_n=1/2$ and $\eta=0$, $(b)$ has $\hat{\omega}_n=2$ and $\eta=0$, plot $(c)$ has $s=\alpha=0$, and $(d)$ has $s=2$ and $\alpha=1/2$.}
    \label{fig:delta-plot}
\end{figure}
As hinted at in the previous section, it is not possible to make a general statement about the effect of negative triangularity on \AE{}; its possible benefit depends strongly on other parameters describing the equilibrium. We can however find trends, and in order to do so we define the following fraction
\begin{equation}
    \Delta = \frac{\widehat{A}(\delta = -0.1)}{\widehat{A}(\delta = +0.1)},
\end{equation}
where $\delta = \pm 0.1$ is chosen to represent a typical experimentally realizable range of parameters. This fraction can be interpreted as the factor by which the \AE{} changes upon switching from positive to negative triangularity, where $\Delta <1$ implies a reduction in \AE{}. We present an investigation of $\Delta$ and its dependencies in Fig. \ref{fig:delta-plot}. We see two clear trends that seem to be robust for tokamaks with $\kappa > 1$. Firstly, as noted in the previous sections, in plots $(a)$ and $(b)$ we see that negative triangularity tends to be especially stabilising for configurations with significant positive shear. Similar conclusions were made by \citet{merlo2023interplay}, who found that the turbulent energy flux in gyrokinetic simulations follows the same trend for TEM-driven turbulence: only for sufficiently high positive shear is a decrease in energy flux found at negative triangularity. Increasing $\alpha$ tends to push the $\Delta = 1$ line (in the plot this is the $\log \Delta = 0$ line) to even higher values of shear, implying that a significant pressure gradient may make negative triangularity less desirable. Secondly, in plots $(c)$ and $(d)$ we note that negative triangularity can be beneficial in situations where the gradient is small, such as in the core. The dependence on $\eta$ is non-trivial; at small density gradients a nonzero value of $\eta$ can make negative triangularity beneficial. As in the previous sections, the results here depend on the Miller vector and are not meant to serve as a quantitative measure for core and edge transport. However, we have found that the presented trends tend to be robust as long as $\kappa > 1$ and thus do have \emph{qualitative} value. We finally note that a more comprehensive model of the effect of negative triangularity should likely take collisions, impurities, and global effects into account \citep{merlo2019turbulent,Merlo2021NonlocalPlasmas}. \par
From these results we infer that negative triangularity is expected to be especially beneficial in the core of the plasma, where gradients are necessarily small. It is not clear if the benefit extends to the edge: only with significant positive shear does negative triangularity become beneficial here as well. One should also keep in mind that $\Delta$ measures the effect of going to negative triangularity while keeping all other parameters fixed. A more complete investigation would, for example, compare experimental equilibria with positive and negative triangularity, or use a global MHD-equilibrium code to find consistent profiles. We do not attempt such an investigation here, but we note that our mathematical framework would readily allow for such a comparison. We finally remark that the above results may seem counter-intuitive as negative triangularity is often thought to automatically imply TEM stabilisation, since the bounce points of most trapped particles reside on the inboard side of the torus, where the magnetic curvature should be favourable. Consequently, it is often argued that the bounce-averaged drift is such that TEMs are stabilised. Upon calculation of \eqref{eq:precession-freq-general}, we find no such stabilisation however, as explained further in Appendix \ref{sec:appendix-tri}.
\subsection{Gradient-threshold like behaviour}
\begin{figure}
    \centering
    \includegraphics[width=\textwidth]{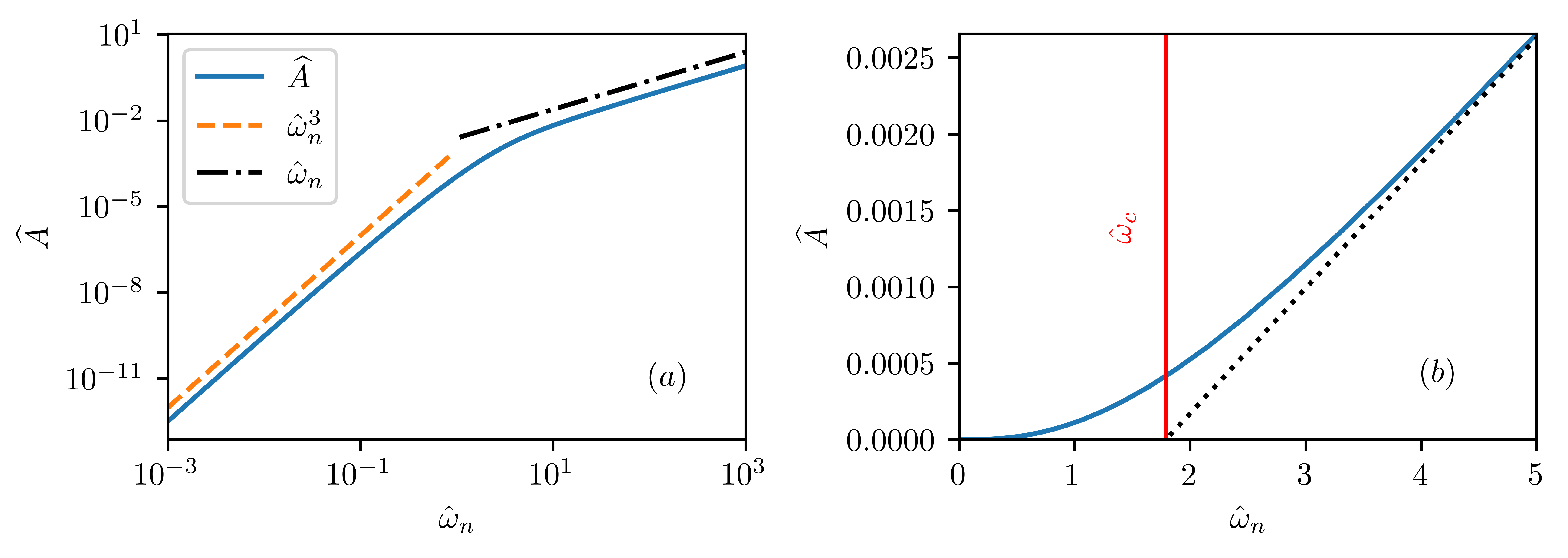}
    \caption{Example of dependence of \AE{} on gradient strength. Two scalings are found in plot $(a)$. In plot $(b)$ we define a gradient threshold by fitting a straight line to the strongly driven regime, and finding its $\hat{\omega}_n$ interception with the abscissa.}
    \label{fig:crit-grad-example}
\end{figure}
Our next step is to investigate the dependence of \AE{} on the gradient strength $\hat{\omega}_n$. From Eq. \eqref{eq:AE-new-coords}, one can show that there are two distinct scalings \citep{Mackenbach2023AvailableTransport}. In a strongly driven regime, one finds that the \AE{} scales linearly with the gradient strength $\hat{\omega}_n$. For a weakly driven regime one can expand around small $\hat{\omega}_n$, and one finds that the \AE{} scales with the gradient strength as $A \propto \hat{\omega}_n^{3}$,
\begin{equation}
    \widehat{A} \propto \begin{cases}
    \hat{\omega}_n   & \text{if } |\hat{\omega}_n | \gg 1, \\
    \hat{\omega}_n^3 & \text{if } |\hat{\omega}_n | \ll 1.
\end{cases}
\end{equation}
These scalings are reminiscent of gradient-threshold (or critical gradient) type behaviour \citep{Dimits2000ComparisonsSimulations}. Gradient thresholds are signified by a sudden decrease in energy flux when decreasing the gradient below some threshold value. The aforementioned scaling behaviour of the \AE{} is displayed in Fig. \ref{fig:crit-grad-example} which similarly shows a rapid decrease below some threshold value. In plot $(b)$ we estimate a critical threshold-like quantity from \AE{}, by fitting a straight line to the strongly driven regime, i.e. we find the best-fit parameters $a_0$ and $a_1$ in the formula
\begin{equation}
    \widehat{A} = a_0 + a_1 \hat{\omega}_n,
\end{equation}
with $\hat{\omega}_n \gg 1$. The gradient threshold, denoted by $\hat{\omega}_c$, is then defined as the interception with the abscissa, hence 
\begin{equation}
    \hat{\omega}_c \equiv - \frac{a_0}{a_1}.
    \label{eq:crit-grad}
\end{equation}
One could, of course, use different definitions for $\hat{\omega}_c$, e.g. one could define the intersection point between the two straight lines on the log-log plot of Fig. \ref{fig:crit-grad-example} as $\hat{\omega}_c$. However, we have found that the definition of Eq. \eqref{eq:crit-grad} has several benefits: it is computationally cheaper, less prone to numerical noise, and seems to behave more smoothly. Other attempted definitions show the same trends. \par
\begin{figure}
    \centering
    \includegraphics[width=\textwidth]{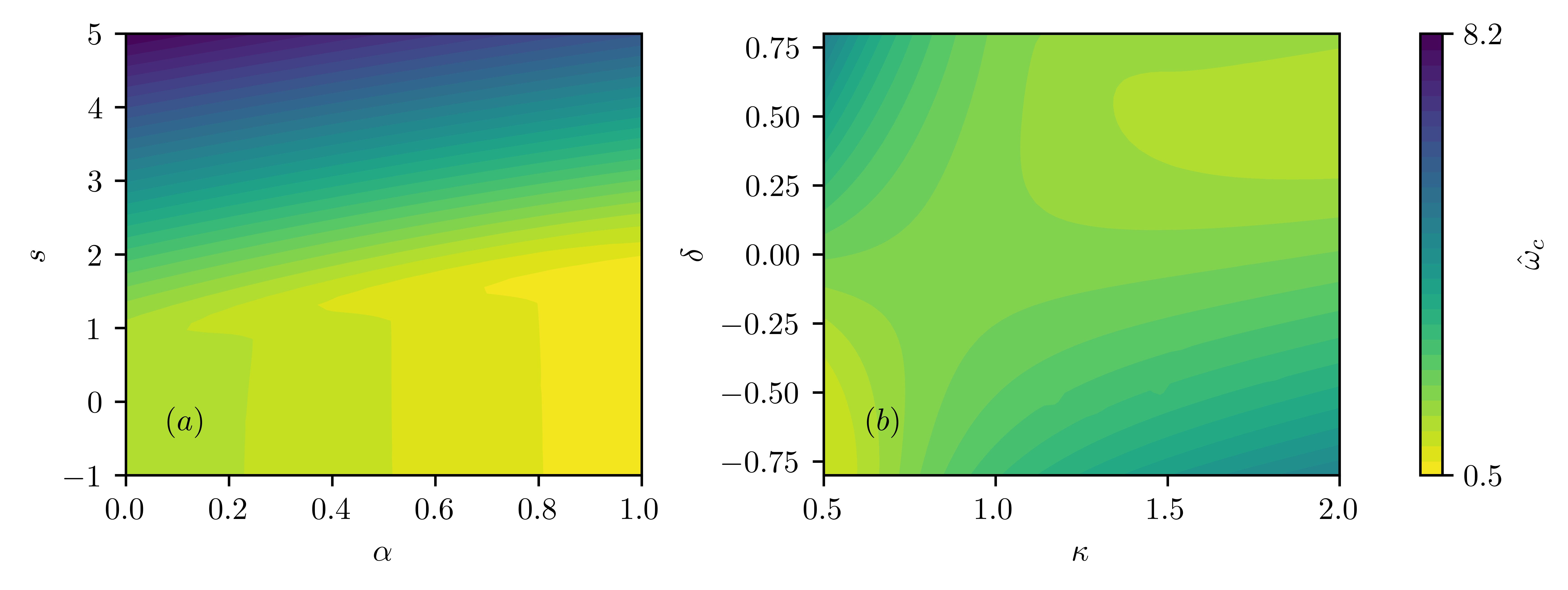}
    \caption{The gradient threshold as a function of equilibrium parameters. In plot $(a)$ the Miller vector $[\epsilon,\kappa,\delta, s_\kappa, s_\delta, \p_r R_0, q, s, \alpha]$ is $[1/3,3/2,1/2,0,0,0,2,s,\alpha]$, and plot $(b)$ has  $[1/3,\kappa,\delta,0,0,0,2,1,0]$. In all plots $\eta = 0$.}
    \label{fig:critgrad}
\end{figure}

 
We illustrate how $\hat{\omega}_c$ varies as a function of various equilibrium parameters in Fig. \ref{fig:critgrad}. Note that subplot $(a)$ in Fig. \ref{fig:critgrad} has the same Miller vector as Fig. \ref{fig:scan-salpha} $(a)$, and subplot $(b)$ in Fig. \ref{fig:critgrad} has the same Miller vector as Fig. \ref{fig:scan-geom} $(a)$. Focussing on plot $(a)$, an interesting trend is that increasing shear tends to increase $\hat{\omega}_c$ linearly, and $\hat{\omega}_c$ tends to plateau for low shear to some value. This is similar to the findings of \cite{jenko2001critical}, though their investigation focusses on electron-temperature gradient turbulence. It is also interesting to note that, in addition to the reduction in \AE{} in the negative-triangularity configuration, it also benefits from a high critical gradient, which is in line with the findings of \cite{Merlo2015InvestigatingTransport}. This effect becomes even more pronounced as one increases the shear, which furthermore reduces the \AE{} in the negative triangularity configuration. This implies that negative triangularity may be beneficial in a different sense: since the critical gradient estimated from \AE{} is higher in negative triangularity geometries, the profiles may be able to sustain much higher gradients and thus higher core density/temperature. 
\subsection{Tokamak optimisation}
In this section we aim to find \AE{}-optimised tokamaks for a certain set of equilibrium parameters, at fixed gradients ($\hat{\omega}_n=1$ and $\eta$=0). To this end, we choose to optimise over $\kappa$ and $\delta$ while keeping all other parameters fixed. In order to find somewhat realistic solutions, we restrict ourselves to a bounded optimisation space, namely
\begin{equation}
    \kappa \in (1/2,2), \qquad \delta \in (-1/2,1/2).
\end{equation}
The SHGO algorithm from \cite{Endres2018AOptimisation} is ideally suited for finding the global minimum in this low-dimensional bounded parameter space and is also available in \texttt{scipy}. Finally, we shall vary magnetic shear and $\alpha$, and investigate its effect on the global minimum found. \par
\begin{figure}
    \centering
    \includegraphics[width=\textwidth]{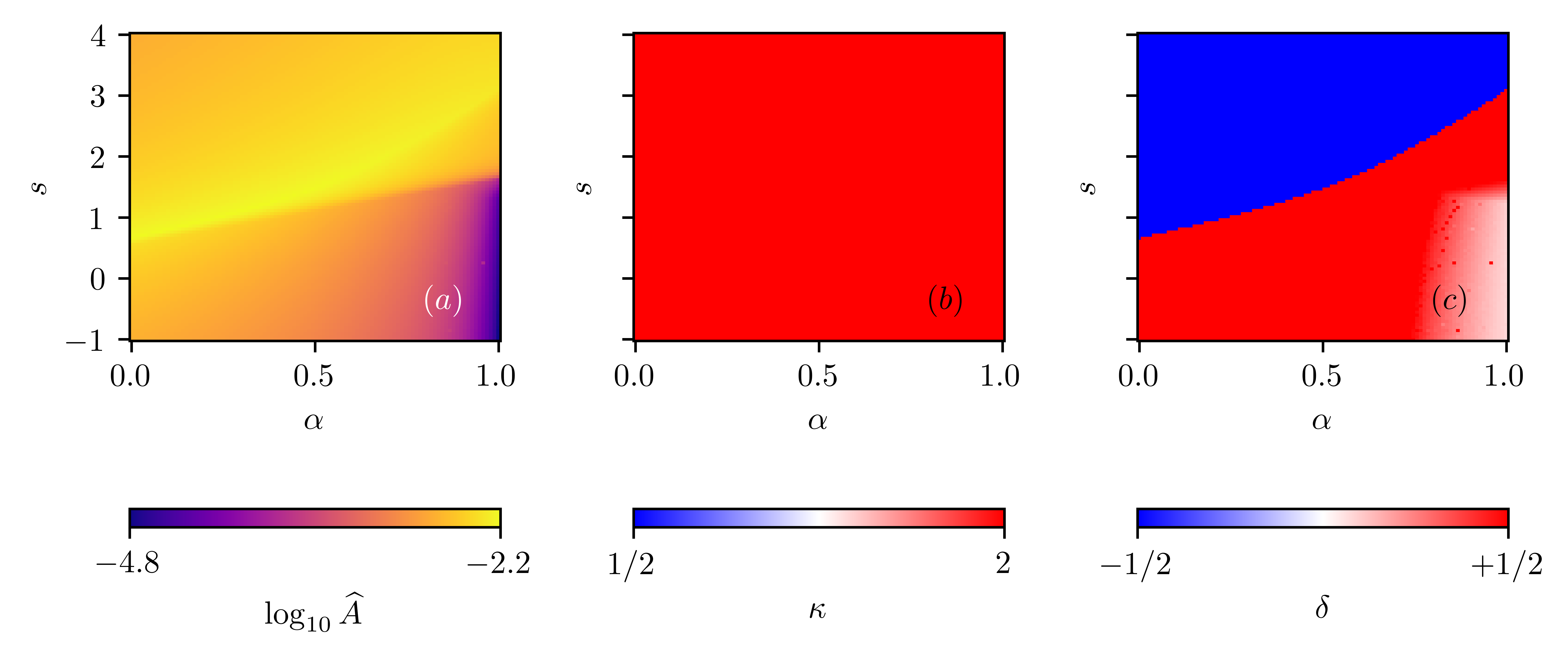}
    \caption{Global \AE{}-minimizing solutions as a function of $s$ and $\alpha$. Plot $(a)$ showcases the \AE{} of the optimal solution, $(b)$ displays the elongation, and $(c)$ shows the triangularity. Generated with a Miller vector $[\epsilon,\kappa,\delta, s_\kappa, s_\delta, \p_r R_0, q, s, \alpha]$ of $[1/3,\kappa,\delta,0,0,-1/2,3,s,\alpha]$. In all plots $\hat{\omega}_n = 1$ and $\eta = 0$.}
    \label{fig:optimisation-results}
\end{figure}

 
The results are displayed in Fig. \ref{fig:optimisation-results}, where the optimal values of \AE{}, $\kappa$, and $\delta$ values are displayed as a function of $s$ and $\alpha$. For a visual aid of the shape of the cross sections, we refer to Fig. \ref{fig:Miller reference shapes}. It can be seen that both the optimal triangularity and elongation tend to be in the corners of the optimisation domain, and hence one should expect that these results are strongly dependent on this domain. Firstly, we see that vertically elongated tokamaks tend to be beneficial for all parameters considered here. It is furthermore interesting to note that the negative triangularity solution tends to be optimal whenever there is significant shear and the pressure gradient is not too large, which is consistent with the findings of Section \ref{sec:neg-triang-ben}. 
\par
From this plot, an important conclusion can be drawn: there is no such thing as a single ``optimal'' solution. The global minimum depends sensitively on other equilibrium parameters, such as shear and pressure gradient, which are, in turn, determined by the profiles of the safety factor, density, and temperature. Therefore, if one is interested in finding an \AE{}-optimised tokamak, one should take care when choosing the profiles. One could also choose to let the profiles be part of the optimisation by describing them with some number of free parameters and constraints (e.g. one could use a fixed number of Fourier modes on top of a profile and optimise for the mode amplitudes). In reality, the profiles are themselves set by equilibrium conditions, making a self-consistent optimisation highly non-trivial. A more consistent investigation could perhaps solve this by coupling the current \AE{}-model to a transport solver, which would calculate self-consistent profiles.
\subsection{Existence of solutions with high gradients yet low \AE{}}
In this section, we investigate how this \AE{} model may relate to the suppression of TEMs when the density gradient is increased. To do so, we note several interesting properties that arise as one increases this gradient. First, the normalised pressure gradient $\alpha$ scales linearly with the density gradient (assuming a constant ratio of the poloidal magnetic field pressure to the thermal pressure, which e.g. occurs if one is operating at a fixed $\beta$-limit). The shear depends on the pressure gradient, as such a gradient drives the bootstrap current, which in turn changes the rotational transform profile. The bootstrap current density has an off-axis maximum in realistic scenarios, and such an off-axis maximum can locally lower the shear. This is most readily seen by inspecting the expression for shear in a large-aspect-ratio, circular tokamak, which depends on the current density profile $j(r)$ as
\begin{equation}
    s(r) = 2 \left( 1 - \frac{j(r)}{\overline{\jmath}(r)} \right); \qquad \overline{\jmath}(r) = \frac{2}{r^2} \int_0^r x j(x) ~ \mathrm{d} x,
    \label{eq:shear-eq}
\end{equation}
where $\overline{\jmath}$ measures the average current density inside the radius $r$. From this expression, it is clear that for current density profiles that peak at $r=0$, the shear is always positive. An off-axis maximum, supplied by the bootstrap current, can cause a locally lower shear. Hence, as one raises $\hat{\omega}_n$ one simultaneously increases $\alpha$ and decreases $s$. To estimate the magnitude of the effect of the bootstrap current on the shear, we note that the bootstrap current is proportional to the density and temperature gradients, and thus to the pressure gradient 
\begin{equation}
    j_b \approx j_{b,0}\alpha(r).
\end{equation}
This is an approximation since the different transport coefficients relating the bootstrap current to the various gradients are not identical \citep{Helander2005CollisionalPlasmas}, but we ignore this minor complication.  We furthermore write the total current density as $j = j_b + j_e$, where $j_e$ is the equilibrium current, and assume $j_b \ll j_e = j_{e,0} \hat{\jmath}(r) $. To first order in the smallness of the bootstrap current, \eqref{eq:shear-eq} then gives
\begin{equation}
    s \approx  2 - \frac{r^2 \hat{\jmath}(r)}{\int_0^\rho  x\hat{\jmath}(x) \mathrm{d}x} \left( 1 + \frac{j_{b,0}}{j_{e,0}} \left[\frac{\alpha(r)}{\hat{\jmath}(r)} - \frac{\int_0^r x \alpha(x) \mathrm{d} x}{\int_0^r x \hat{\jmath}(x) \mathrm{d} x} \right] \right) .
\end{equation}
Finally, following  \cite{miyamoto2005plasma} we estimate the ratio $j_{b,0}/j_{e,0}$ as 
\begin{equation}
    \frac{j_{b,0}}{j_{e,0}} \approx 0.3 \langle \beta_p \rangle \sqrt{\epsilon},
\end{equation}
where $\beta_p$ is the local ratio of the thermal pressure over the poloidal magnetic field pressure, and the angular brackets denote a volume average. We shall take $j_{b,0}/j_{e,0}$ to be on the order of $10 \%$, implying that the shear may change as $\mathrm{d} s /\mathrm{d} \alpha  \sim s/10$. Finally, one can relate the pressure gradient to $\hat{\omega}_n$ as
\begin{equation}
    \alpha = \epsilon \beta_p \left( 1 + \eta + \eta_i \right)\hat{\omega}_n ,
\end{equation}
where $\eta_i=\partial_r \ln T_i / \partial_r \ln n$, with $T_i(r)$ being the ion temperature. We assume that the factor $\epsilon \beta_p (1 + \eta + \eta_i) \sim 0.1$, so that $\mathrm{d} \hat{\omega}_n / \mathrm{d} \alpha \sim 10$.
\par  
\begin{figure}
\centering
\begin{subfigure}[b]{1.0\textwidth}
   \includegraphics[width=\textwidth]{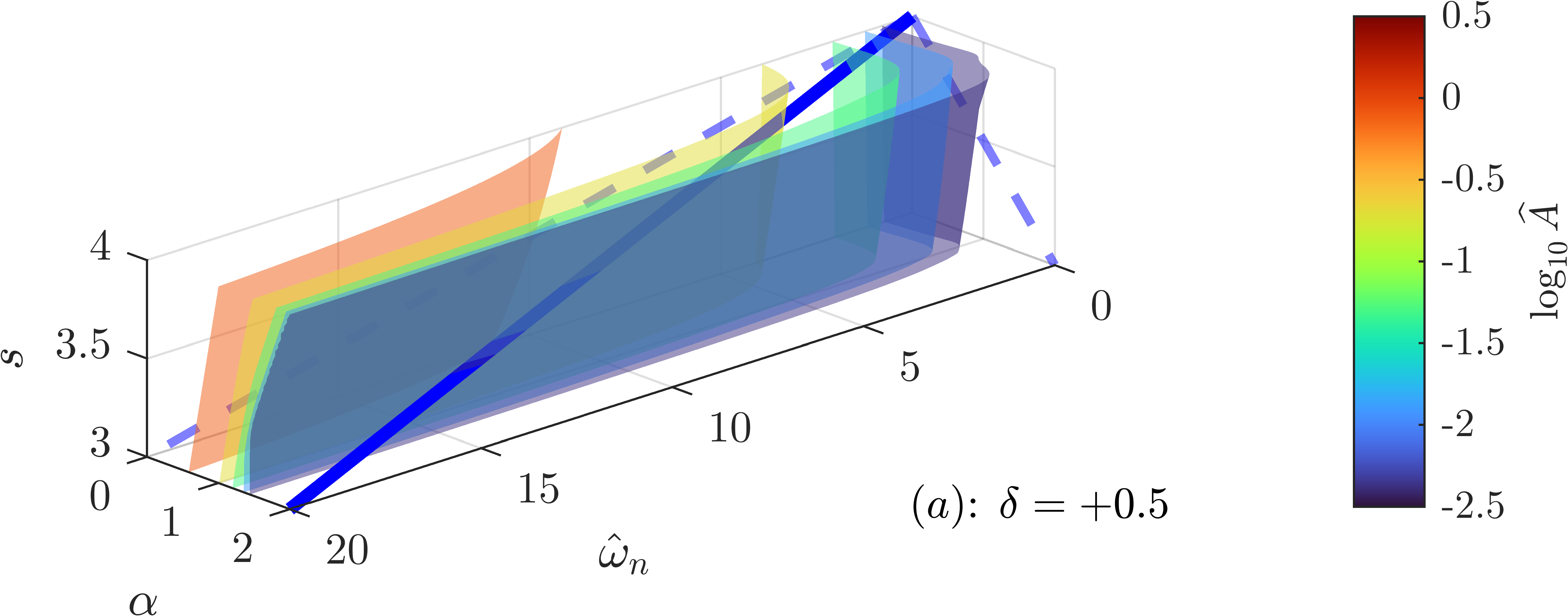}
   \label{fig:Ng1} 
\end{subfigure}

\begin{subfigure}[b]{1.0\textwidth}
    \includegraphics[width=\textwidth]{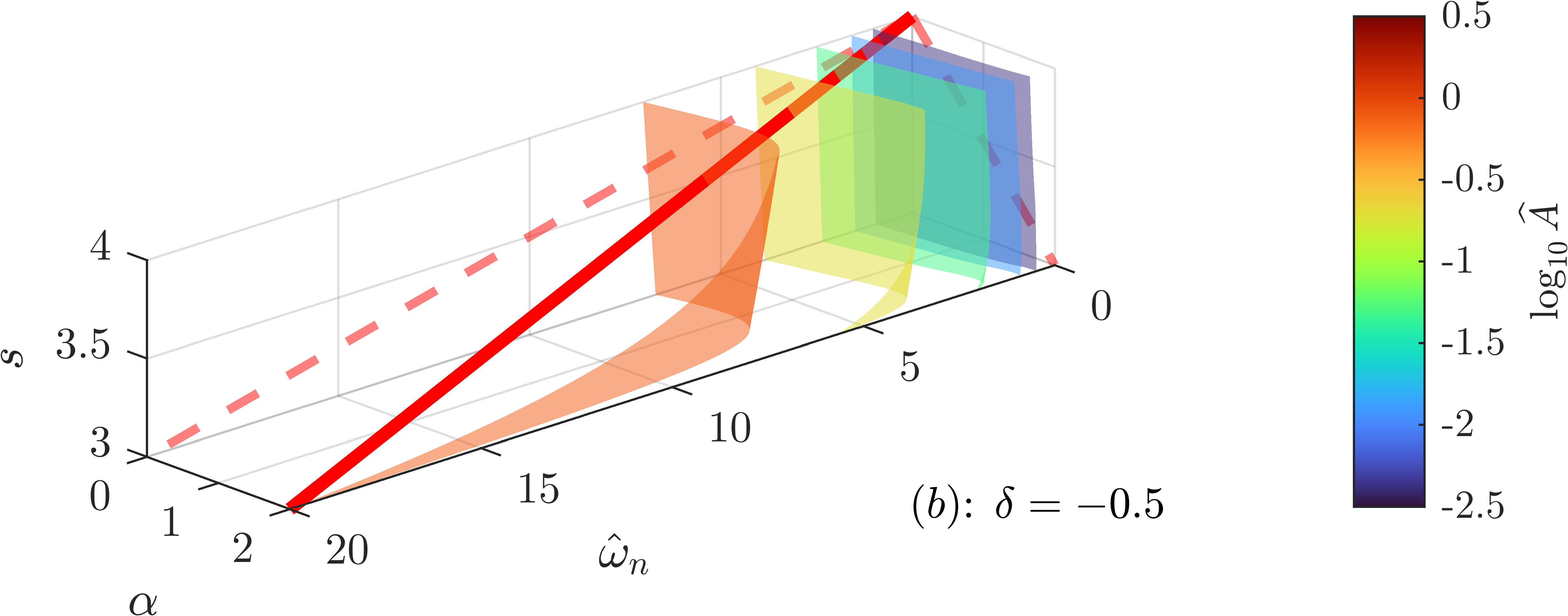}
   \label{fig:Ng2}
\end{subfigure}

\begin{subfigure}[b]{1.0\textwidth}
    \includegraphics[width=\textwidth]{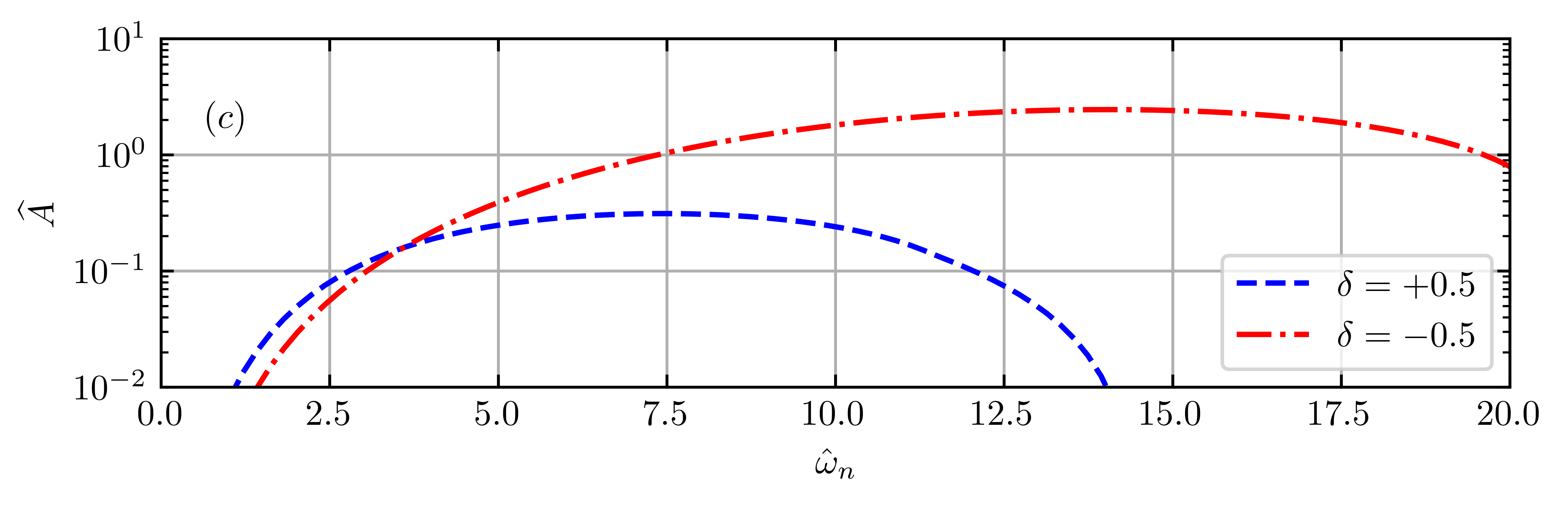}
   \label{fig:Ng3}
\end{subfigure}
\caption[three plots]{subfigures $(a)$ and $(b)$ showcase iso-contours of the \AE{} as a function of $(\hat{\omega}_n,\alpha,s)$, where subfigures $(a)$ and $(b)$ have positive and negative triangularity respectively. In both $(a)$ and $(b)$ a straight line is drawn which has increasing $\alpha$ and decreasing $s$ with increasing $\hat{\omega}_n$, and the projection of the line onto the grid-planes is shown as a dashed line. In subfigure $(c)$, the \AE{} along both the blue line of subfigure $(a)$ and the red line of subfigure $(b)$ is plotted as a function $\hat{\omega}_n$. These plots were generated with a Miller vector $[\epsilon,\kappa,\delta, s_\kappa, s_\delta, \p_r R_0, q, s, \alpha]$ of $ [1/3,3/2,\delta, 1/2, \delta/\sqrt{1-\delta^2}, -1/2, 3, s, \alpha]$, and $\eta = 0$.}
\label{fig:isocontour-AE}
\end{figure}
We illustrate the competing effects of the density gradient, pressure gradient, and shear in Fig. \ref{fig:isocontour-AE}. In subfigures $(a)$ and $(b)$ we see various iso-contours of the \AE{} in $(\hat{\omega}_n,s,\alpha)$-space, where $(a)$ has positive triangularity and $(b)$ has negative triangularity. It is especially interesting to note that in subfigure $(a)$ there are paths in parameter space in which  $\hat{\omega}_n$ increases but the \AE{} \textit{decreases}. These paths generally require that, as the density gradient increases, the pressure gradient should also increase and the shear should decrease. As we have argued, these trends are indeed found in tokamak discharges. One such path is indicated in subplot $(a)$ as a blue line. Importantly, the blue line has
\begin{equation*}
    s = 4 \left(1  - \frac{\alpha}{8} \right), \qquad \hat{\omega}_n = 10 \cdot \alpha
\end{equation*}
which is the right order of magnitude for both $\mathrm{d}s/\mathrm{d}\alpha$ and $\mathrm{d}\hat{\omega}_n/\mathrm{d}\alpha$. Subfigure $(b)$ exhibits drastically different features. Planes of constant \AE{} tend to lie parallel to planes of constant $\hat{\omega}_n$, indicating that not much stabilisation is possible by changing the shear or the pressure gradient: the \AE{} rises when $\hat{\omega}_n$ is increased. In subfigure $(b)$, we again plot a line along the direction of increasing $\alpha$ and decreasing magnetic shear in red. Finally, note that for $s_\delta$ we have used the estimate from \cite{Miller1998NoncircularModel}, $s_\delta \approx \delta/\sqrt{1 - \delta^2}$. \par 
In subfigure $(c)$ we display the \AE{} along the blue and red lines given in subfigures $(a)$ and $(b)$ as a function of the density gradient. Note that the positive-triangularity case exhibits a distinct maximum, with low \AE{} both to the left and right of the peak. One could interpret the existence of the latter as two distinct low-transport regimes; one with low gradients, and one with high gradients (which also has decreased magnetic shear and increased $\alpha$). It is furthermore interesting to note that the negative-triangularity tokamak rises to far higher values in terms of \AE{} and does not seem to drop back down to low levels along the chosen domain. Hence one could perhaps conclude that reaching a low-transport state with high gradients is not feasible in a negative-triangularity discharge. This is in line with findings of \cite{Saarelma2021BallooningTokamak} and \cite{nelson2022prospects}, where the H-mode was found to be inaccessible in negative-triangularity tokamaks on basis of the ballooning instability, though the physical reason is of course different. This rise in \AE{} in negative triangularity is perhaps unsurprising given that we have found that negative triangularity is stabilising in cases with significant positive shear, a weak pressure gradient, and a slight density gradient, exemplified in Figs. \ref{fig:delta-plot} and \ref{fig:optimisation-results}. Since, along the chosen path shear decreases and $\alpha$ increases with increasing density gradient, which is opposite to what is stabilising for negative-triangularity tokamaks, we see a sharp increase in \AE{}.  It may be feasible, however, to have a significant reduction in transport by tailoring the $q$-profile in such a way that negative triangularity becomes favorable, which likely implies significant positive shear. With such a reduction in \AE{}, one could perhaps enjoy much improved transport whilst staying in an L-mode like regime. The parameters described in \cite{Marinoni2019H-modeDIII-D} do seem to meet such requirements, especially near the edge where the reduction in transport seems greatest as compared to the positive triangularity case. \par 
A more comprehensive investigation, which shall be undertaken in a future publication, would self-consistently calculate the bootstrap current which would give precise paths in $(\hat{\omega}_n,\alpha,s)$-space. However, given the nature of the iso-contours in this three-dimensional space, we expect the observed trends to be robust, as long as the path has the correct general dependencies (i.e. decreasing shear and increasing $\alpha$ with increasing density gradient).

\section{Conclusions}
We have shown that it is possible to simplify the analytical expression for the \AE{} of trapped electrons in the case of an omnigenous system, which speeds up calculations. If one furthermore employs an analytical local solution to the Grad-Shafranov equation, explicit expression of various quantities needed in the calculation of the \AE{} (e.g., bounce-averaged drifts, bounce times) can be found as in \cite{Roach1995TrappedTokamaks}. Making use of an equilibrium parameterisation proposed by \cite{Miller1998NoncircularModel}, we go on to investigate how \AE{} depends on these equilibrium parameters. Using this set-up, we observe several interesting features of the \AE{}:
\begin{enumerate}
    \item A comparison is made between \AE{} and \textsc{tglf}. We observe a fairly good correlation between energy flux and $A^{3/2}$, indicating that \AE{} can be a useful measure for tokamak transport.
    \item Increasing the magnitude of the magnetic shear or increasing the Shafranov shift tends to be stabilising as indicated by a reduction in the \AE{}, and these trends hold for many different choices of geometry. Especially negative shear reduces the \AE{} substantially for pure density gradients.
    \item Vertical elongation tends to be stablising, as indicated by a reduction in \AE{}. Negative triangularity can be stabilising, particularly in configurations with significant positive shear or small gradients, but not always. 
    \item The \AE{} has different scalings with respect to the gradient strength in weakly and strongly driven regimes. We employ this difference in scaling to estimate a gradient-threshold like quantity, and we find that it has similar behaviours as found in critical-gradient literature; an increase in shear tends to increase this gradient-threshold and negative triangularity benefits from an especially high gradient-threshold.
    \item Using \AE{} for shape-optimisation we show that the optimal solution is strongly dependent on pressure gradients and magnetic shear, implying that the optimisation is sensitive to the density, pressure, and $q$-profiles.
    \item An investigation is presented on how \AE{} varies as the density and pressure gradient increase consistently, while shear decreases. We find that in such scenarios one can find solutions with large gradients yet low \AE{}. Such solutions tend to exist for positive triangularity tokamaks but not for negative triangularity tokamaks.
\end{enumerate}
\par
The results suggest that various observed trends regarding turbulent transport in tokamaks may partly be understood in terms of \AE{}, which has a simple physical interpretation and is cheap to compute. The analytical framework can readily be extended to account for an equilibrium model which allows for other shaping and plasma parameters such as plasma rotation \citep{Hameiri1983ThePlasmas,Miller1995StabilizationRotation}, squareness \citep{Turnbull1999ImprovedTokamaks}, and up-down asymmetry \citep{Rodrigues2018LocalPlasmas}, though no such investigation is presented here.
\section*{Acknowledgments}
We wish to thank J. Ball, J.M. Duff, R. Wolf, A. Goodman, P. Mulholland, P. Costello, M.J. Pueschel, F. Jenko, M. Barnes, and E. Rodriguez for insightful discussions. This work was partly supported by a grant from the Simons Foundation (560651, PH), and this publication is part of the project “Shaping turbulence—building a framework for turbulence optimisation of fusion reactors,” with Project No. \texttt{OCENW.KLEIN.013} of the research program “NWO Open Competition Domain Science” which is financed by the Dutch Research Council (NWO). This work has been carried out within the framework of the EUROfusion Consortium, funded by the European Union via the Euratom Research and Training Program (Grant Agreement No. 101052200—EUROfusion). Views and opinions expressed are however those of the author(s) only and do not necessarily reflect those of the European Union or the European Commission. Neither the European Union nor the European Commission can be held responsible for them.
\section*{Competing interests}
The authors declare no competing interests.
\appendix
\section{Details of bounce-averaging integrals in the $s$-$\alpha$ limit}
\label{sec:details-derivation-of-s-alpha-integrals}
In the large-aspect ratio limit with circular flux surfaces, we find that 
\begin{subequations}
\begin{eqnarray}
\gamma &=& 1, \\
C_1 &=& -1, \\
C_2 &=& 0, \\
C_3 &=& 1, \\
C_4 &=& 1, \\
\xi &=& 2 \pi , \\
\oint \hat{l}_\theta \hat{B}_{p,s}^{-1} &=& 2 \pi .
\end{eqnarray}
\end{subequations}
The equation for shear simplifies to $\sigma = s - 2 $. Next, we investigate the radial derivatives of the magnetic field components in this limit and find that these become 
\begin{subequations}
\begin{eqnarray}
r \partial_\rho b_p &=& 1 - s + \alpha \cos \theta, \\
r \partial_\rho b &=& \epsilon \left( \frac{\alpha}{2 q^2} - \cos \theta \right).
\end{eqnarray}
\end{subequations}
We express $\lambda$ in terms of the trapping parameter $k^2$, where the deeply trapped particles have $k=0$ and the barely trapped particles have $k=1$. This mapping is given by $\lambda = 1 + \epsilon (1 - 2 k^2)$, so that the magnetic field may be written as
\begin{equation}
    \lambda \hat{B} = 1 + \epsilon (1 - 2 k^2 - \cos \theta).
\end{equation}
We can now express the argument of the bounce-averaging operator of Eq. \eqref{eq:precession-freq-general}, and expand it around the smallness of $\epsilon$. This gives us the leading order result,
\begin{equation}
\begin{aligned}
    \hat{\omega}_\lambda = & \left\langle -\frac{\alpha}{2 q^2} + \cos \theta + 2 (2 k^2 + \cos \theta - 1 )(s - \alpha \cos \theta) \right\rangle_\lambda.
\end{aligned}
\end{equation}
\begin{figure}
    \centering
    \includegraphics[width=0.5\textwidth]{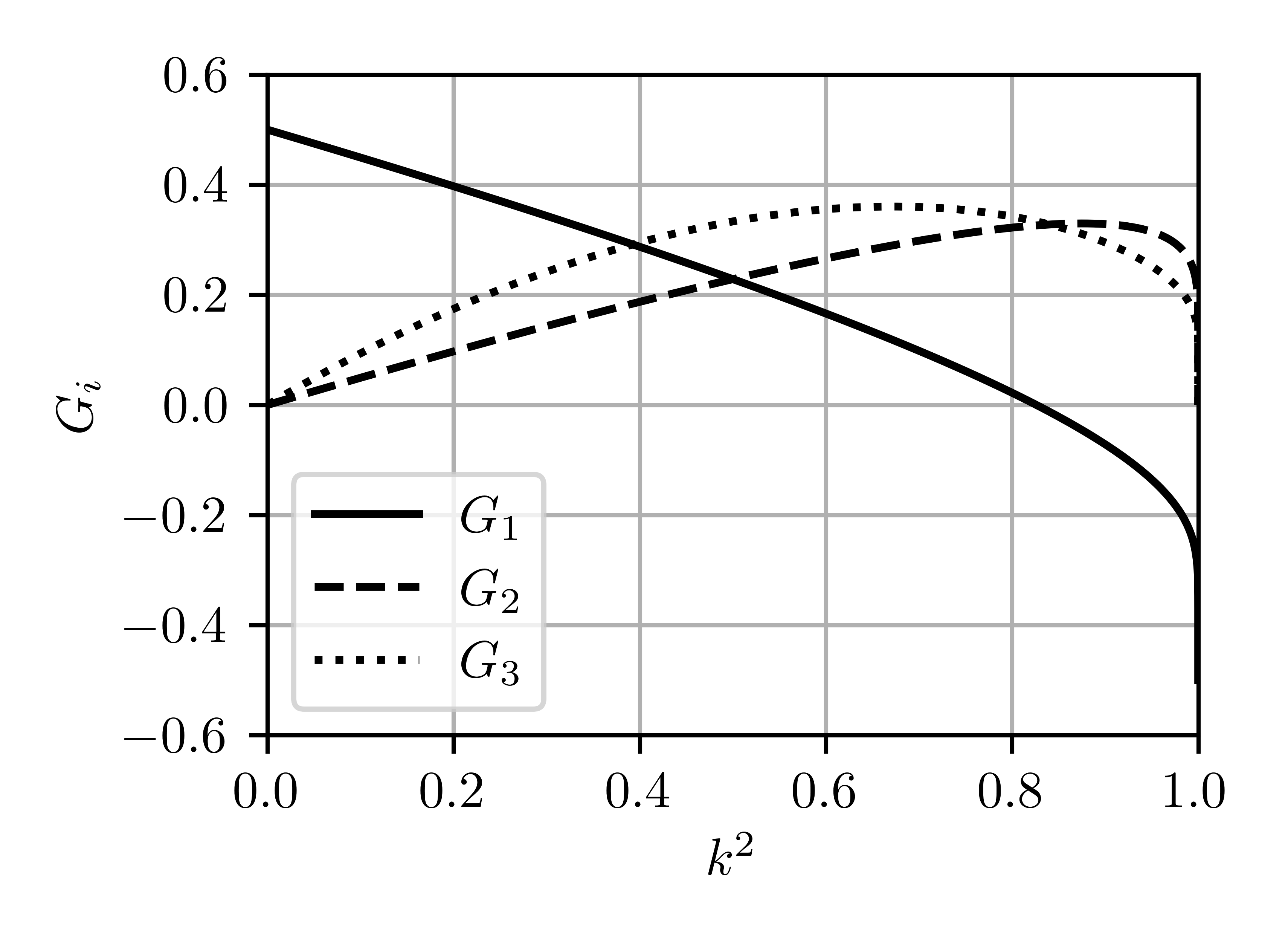}
    \caption{The functions $G_1$, $G_2$, and $G_3$ as a function of the trapping parameter $k^2$.}
    \label{fig: CHM drifts}
\end{figure}
In order to evaluate the integral, we first consider the general problem of evaluating
\begin{equation}
    I = \int_{\rm bounce} \frac{f(\theta) \mathrm{d} \theta }{\sqrt{\epsilon} \sqrt{2k^2 + \cos{\theta} - 1}},
\end{equation}
where the region of integration is set by the region where the argument of the square root is positive. Using the double angle identity $\cos(\theta) = 1 - 2 \sin^2 (\theta / 2)$, and setting $\tilde{\theta} = \theta/2$ gives
\begin{equation}
    I = \sqrt{\frac{2}{\epsilon}} \int_{\rm bounce} \frac{f(2 \tilde{\theta}) \mathrm{d} \tilde{\theta} }{\sqrt{k^2 - \sin^2{\tilde{\theta}}}}.
\end{equation}
Next, one uses the $u$-substitution $\sin\tilde{\theta} = k\sin \vartheta$, which has
\begin{equation}
    \mathrm{d} \tilde{\theta} = \frac{\sqrt{k^2 - k^2 \sin^2 \vartheta}}{\sqrt{1 - k^2 \sin \vartheta}} \mathrm{d} \vartheta,
\end{equation}
so that the integral becomes
\begin{equation}
    I = \sqrt{\frac{2}{\epsilon}} \int_{-\pi/2}^{\pi/2} \frac{f(2 \arcsin[k \sin \vartheta] ) \mathrm{d} \vartheta }{\sqrt{1- k^2\sin^2{\vartheta}}},
\end{equation}
where have recognised the limits of integration satisfy $\vartheta = \pm \pi/2$. The integral is now in standard form, and may be related to elliptic integrals of the first and second kind, depending on the functional form of $f$. For any constant function $f(\theta) = f_0$, one simply has
\begin{equation}
    I = f_0 \sqrt{\frac{2}{\epsilon}} \int_{-\pi/2}^{\pi/2} \frac{\mathrm{d} \vartheta }{\sqrt{1- k^2\sin^2{\vartheta}}} =2 f_0 K(k) \sqrt{\frac{2}{\epsilon}} ,
\end{equation}
where the elliptic integral of the first kind is $K(k) = \int_0^{\pi/2} \mathrm{d} \vartheta/\sqrt{1 - k^2 \sin \vartheta}$. Next, we require the integral with $f(\theta) = \cos \theta = 1 - 2 k^2 \sin^2 \vartheta $. This becomes
\begin{equation}
    I = 2\sqrt{\frac{2}{\epsilon}} \int_{\pi/2}^{\pi/2} \frac{1 - 2 k^2 \sin^2 \vartheta }{\sqrt{1- k^2\sin^2{\vartheta}}}\mathrm{d} \vartheta = 2\sqrt{\frac{2}{\epsilon}} \left( 2 E(k) - K(k) \right),
\end{equation}
where $E(k) = \int_0^{\pi/2} \mathrm{d} \vartheta\sqrt{1 - k^2 \sin \vartheta}$ is the elliptic integral of the first kind. We finally require the integral with $f(\theta) = \cos^2 \theta$, which reduces to
\begin{equation}
    I = \frac{2}{3}\sqrt{\frac{2}{\epsilon}} \left( \left[ 4 - 8k^2 \right] E(k) + \left[4 k^2 - 1\right] K(k) \right).
\end{equation}
The bounce-average of the large-aspect ratio tokamak may now be evaluated, and one finds the result given in \eqref{eq: w CHM}, equivalent to the result of \citet{Connor1983EffectTokamak}. A plot of all these functions may be found in Fig. \ref{fig: CHM drifts}. As a final step, we calculate the dimensionless bounce-time, given Eq. \eqref{eq: dimless bounce-times}. We find that it reduces to
\begin{equation}
    \hat{g}_\epsilon^{1/2} = \frac{\sqrt{2}}{\pi} K(k).
\end{equation}
Inserting the found results into Eq. \eqref{eq:ae-final-with-Cr} gives the result given in Eq. \eqref{eq:s-alpha-exact}

\section{Benchmark of circular tokamak and Miller code and asymptotic limits} \label{sec:appendix-salpha-miller-comparison}
Here we show that the two codes that calculate the \AE{} in both the circular $s$-$\alpha$ tokamak, for which the equation is given in \eqref{eq:s-alpha-exact}, and a Miller tokamak, as given in Eq. \eqref{eq:ae-final}, indeed yield the same results in the limit of a large aspect ratio circular tokamak. For a proper comparison, we set the Miller parameters such that one approaches the $s$-$\alpha$ limit. As such, we choose $\epsilon=10^{-6}$, $q=2$, and all other Miller components of the Miller vector as given in \eqref{eq:miller-vector} are set to zero. There is one numerical parameter of interest in the Miller code, the number of $\theta$ points which are used to evaluate the bounce integrals of Eq. \eqref{eq:precession-freq-general} using a generalised trapezoidal method \citep{mackenbach2023drift}. In the comparison presented here we use $10^3$ equidistant nodes for $\theta$. The integral over the pitch angle is done using quadrature methods. \par
\begin{figure}
    \centering
    \includegraphics[width=\textwidth]{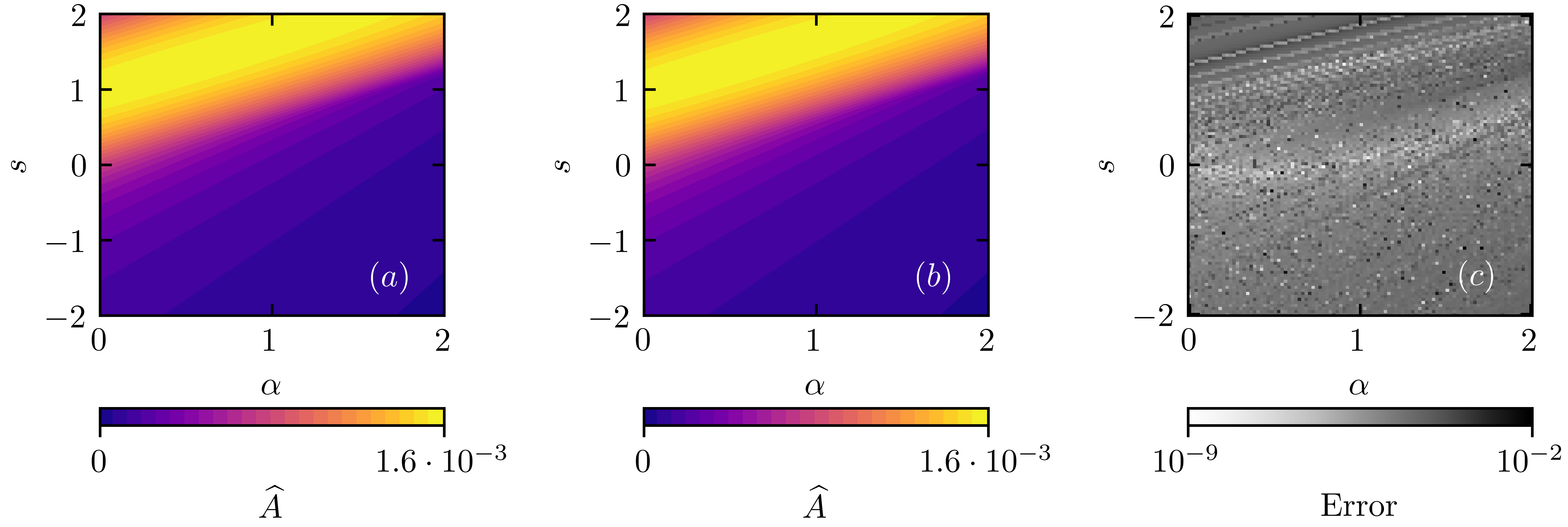}
    \caption{Comparison of the \AE{} as calculated with two different codes. Plot $(a)$ is calculated using a code that calculates \AE{} in the $s$-$\alpha$ limit, and $(b)$ is calculated using the Miller code. The plots are visually indistinguishable. Calculated using $q=2$, $\epsilon = 10^{-6}$, $\hat{\omega}_n = 3$, and $\eta=1$. All other parameters for Miller are set to zero, as required in the limit of the $s$-$\alpha$ tokamak. Plot $(c)$ presents the relative error, where it can be seen that the relative error is very small for large regions of $s$-$\alpha$ space. Note that the colourbar scale in plot $(c)$ is logarithmic.}
    \label{fig:code-comparison}
\end{figure}

 
The comparison is shown in Fig. \ref{fig:code-comparison}. In this figure, three different contour plots are shown; $(a)$ is the available energy as calculated from Eq. \eqref{eq:s-alpha-exact}. Plot $(b)$ shows the result as calculated from Eq. \eqref{eq:ae-final}. Finally, plot $(c)$ shows the relative error between the two codes (more precisely, it is the difference between plot $(a)$ and $(b)$, divided by plot $(a)$). It can be seen that the error is typically quite small, with a maximal value of 1\% and a mean value of $0.004 \%$. If different parameters are chosen (safety factor, density gradient, or $\eta$), the error remains similarly small. \par
All plots presented in the current publication are generated using the same or even more refined numerical parameters as used here, so that we have a high degree of confidence that the presented trends are indeed physical and not numerical. Further convergence checks (increasing the resolution of $\theta$ and adjusting the tolerances of the quadrature methods) do not alter the plots presented in this publication in a visually discernible manner. \par
\begin{figure}
    \centering
    \includegraphics[width=\textwidth]{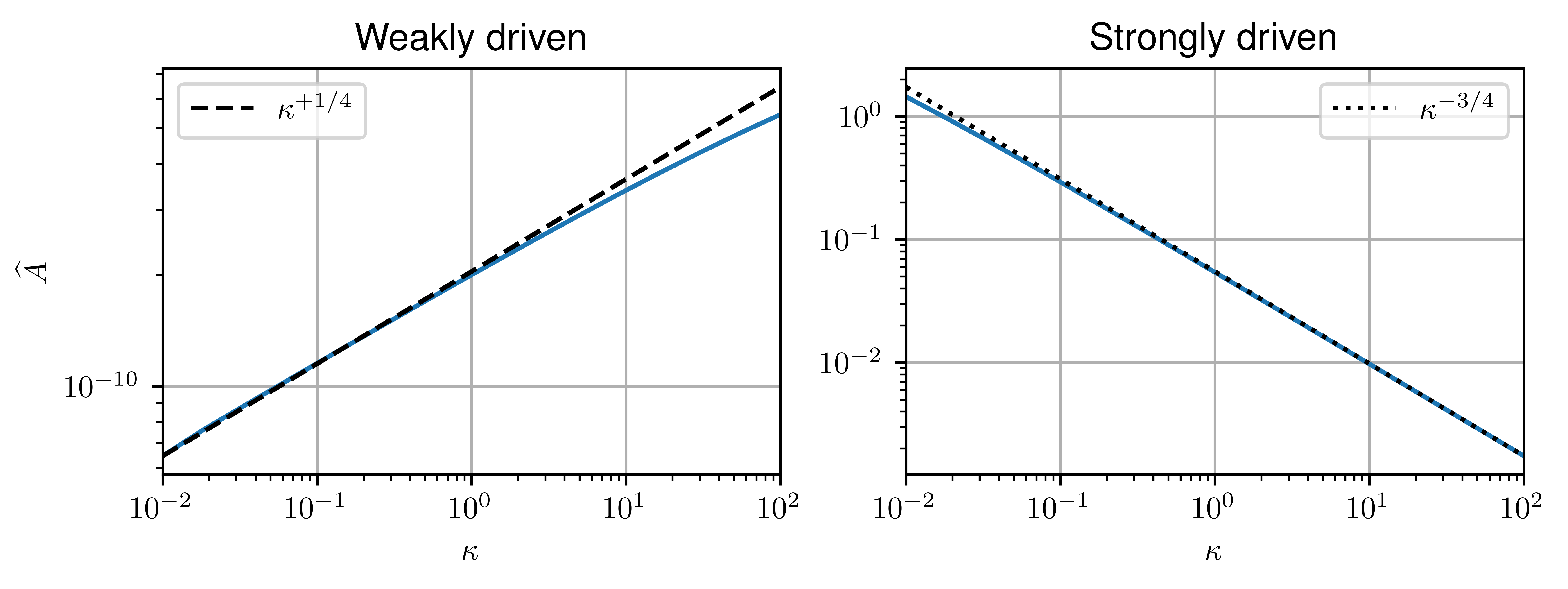}
    \caption{Available energy as a function of the elongation $\kappa$ for a very weak density gradient in the left plot ($\hat{\omega}_n = 1/100$ at $\kappa = 1$), and a very strong density gradient in the right plot ($\hat{\omega}_n = 100$ at $\kappa = 1$).}
    \label{fig:elongation-scaling}
\end{figure}

 
As an additional check, we highlight that a recent publication has evaluated the \AE{} of trapped electrons in quasi-symmetric systems (which includes tokamaks) in two asymptotic limits: those of a very strong and a very weak density gradient \citep{rodriguez2023trapped}. It was found that the \AE{} scales with elongation as $\widehat{A} \propto \kappa^{1/4}$ if the density gradient is sufficiently small, and $\widehat{A} \propto \kappa^{-3/4}$ if the density gradient is sufficiently large. Importantly, this analysis assumed fixed $\partial n / \partial r_{\rm eff}$ and $r_{\rm eff} / R_0$, instead of fixed $\partial_r n$ and $r/R_0$. If one properly accounts for this different definition of the radial coordinate, we find that the code reproduces the correct scaling behaviours, as may be seen in Fig. \ref{fig:elongation-scaling}.
We also note that in the aforementioned investigation it was found that at zero shear, more negative triangularity is found to increase the \AE{} if the gradient is sufficiently strong and $\kappa > 1$. If the density gradient is sufficiently strong and $\kappa < 1$, the \AE{} decreases with more negative triangularity. These trends are reproduced and can be found in Fig. \ref{fig:scan-geom}, subplot $(d)$.
\section{Negative triangularity and trapped particle precession} \label{sec:appendix-tri}
\begin{figure}
    \centering
    \includegraphics[width=\textwidth]{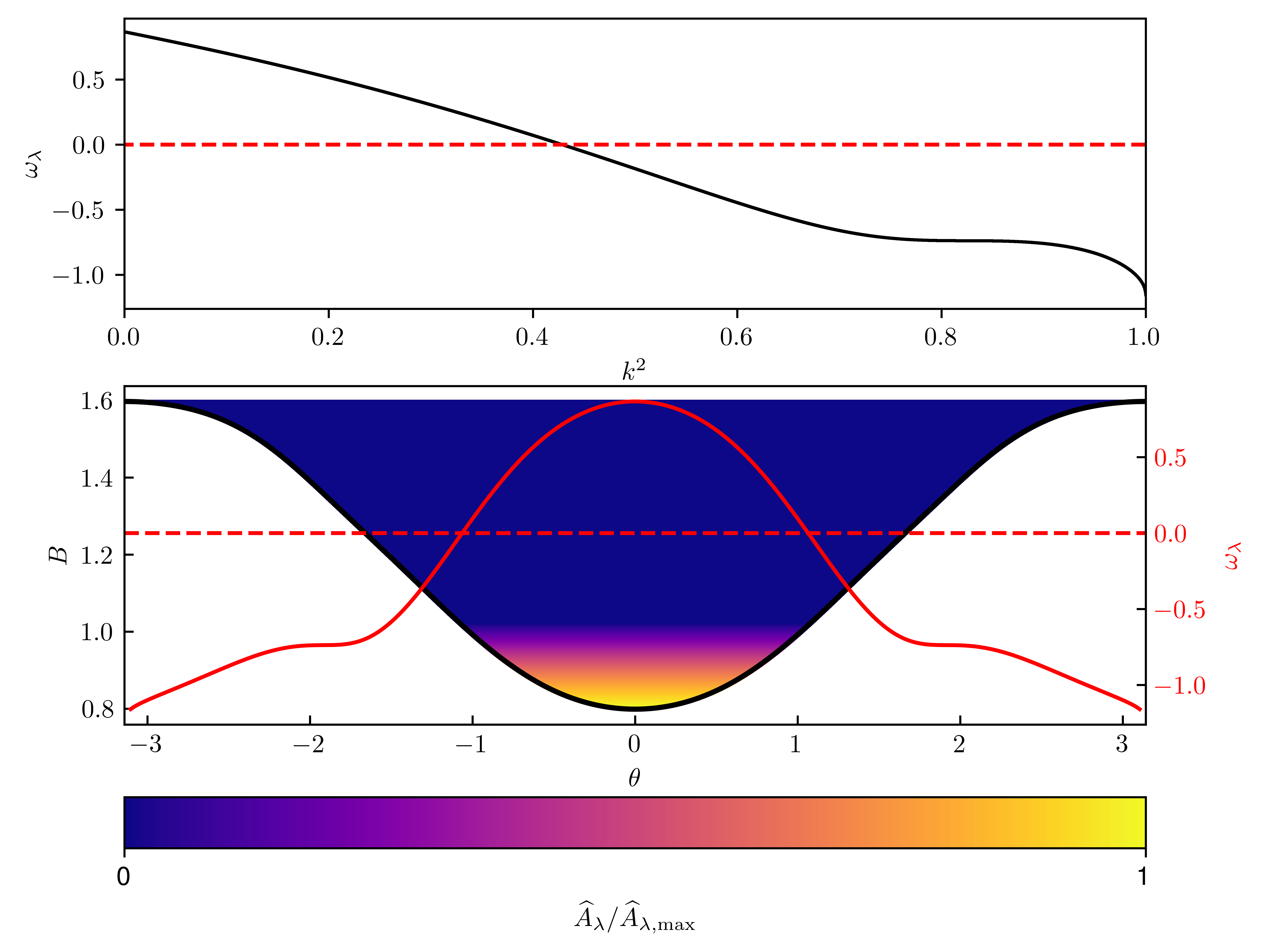}
    \caption{The precession frequency and \AE{} distribution for a positive triangularity tokamak}
    \label{fig:pos-triangularity-tok}
\end{figure}
In this section, we investigate the difference in trapped particle orbits in positive and negative triangularity tokamaks. To this end, we investigate the dependence of Eq. \eqref{eq:precession-freq-general} on $\delta$, and we set the other components of the Miller vector equal to $[\epsilon,\kappa,\delta, s_\kappa, s_\delta, \p_r R_0, q, s, \alpha] = [1/3,2,\delta,0,0,0,2,0,0]$. \par  
\begin{figure}
    \centering
    \includegraphics[width=\textwidth]{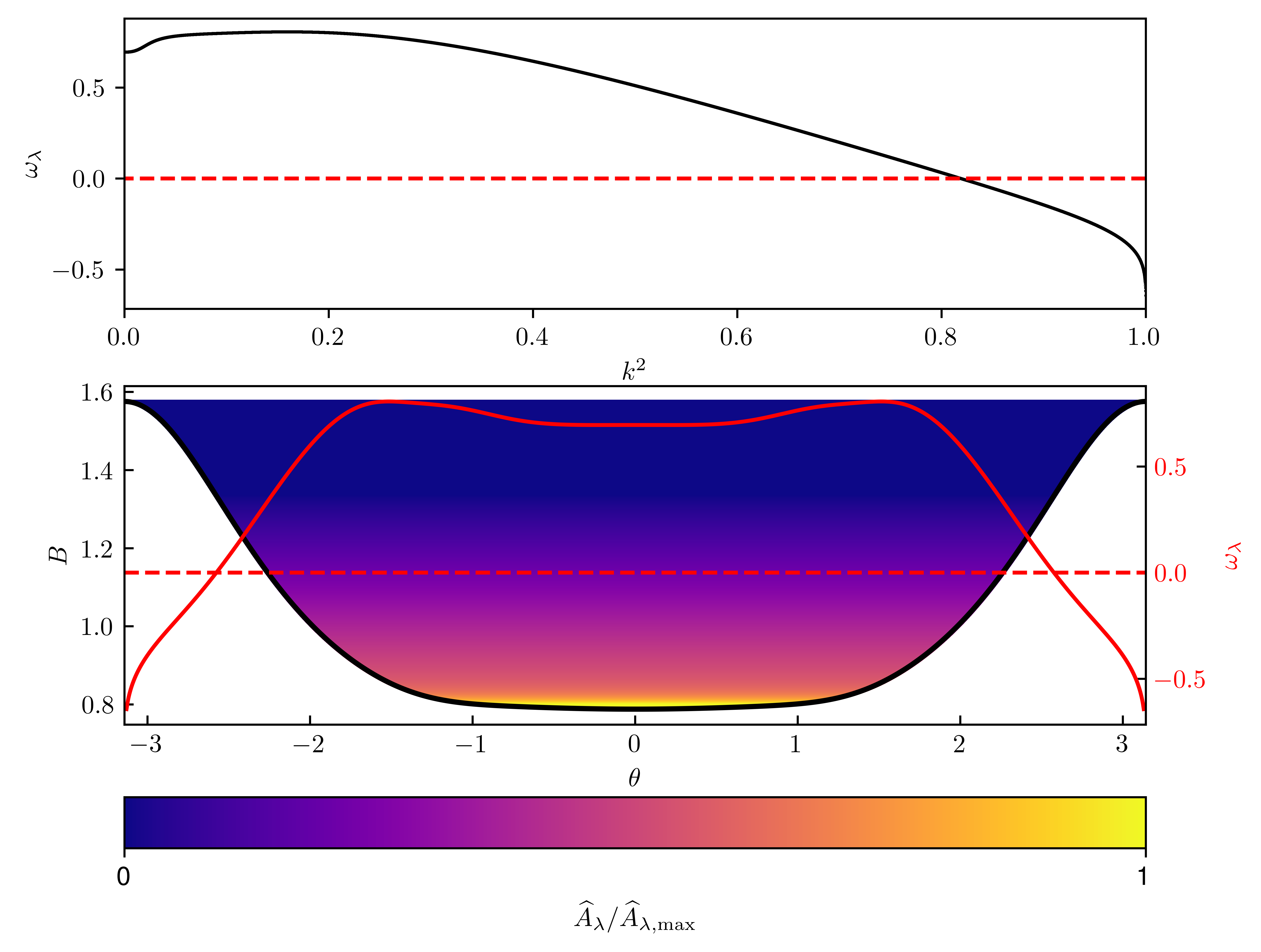}
    \caption{The precession frequency and \AE{} distribution for a negative triangularity tokamak}
    \label{fig:neg-triangularity-tok}
\end{figure}
The result for a positive triangularity tokamak ($\delta = 0.5$) is plotted in Fig. \ref{fig:pos-triangularity-tok}, where we have plotted $\omega_\lambda$ as a function of its bounce points $\theta$, which satisfy
\begin{equation}
    1 - \lambda \hat{B}(\theta) = 0.
\end{equation}
Furthermore, we have shown the \AE{} per $\lambda$, called $\widehat{A}_\lambda$, which is the integrand of Eq. \eqref{eq:ae-final}. This is done by coloring a line of constant $\lambda$ (which corresponds to constant $B$) according to its $A_\lambda$. Finally, we also display $\omega_\lambda$ as a function of the trapping parameter $k^2$ which maps $\lambda \mapsto [0,1]$ according to
\begin{equation}
    k^2 = \frac{\hat{B}_\mathrm{max}- \lambda \hat{B}_\mathrm{max} \hat{B}_\mathrm{min}}{\hat{B}_\mathrm{max}- \hat{B}_\mathrm{min}},
\end{equation}
where the subscripts $\mathrm{max}$ and $\mathrm{min}$ refer to the maximal and minimal values of the functions respectively. With this convention, $k^2 = 0$ corresponds to the most deeply trapped particles and $k^2=1$ to the most shallowly trapped particles. We have furthermore included a red dashed line, which delineates where $\omega_\lambda$ changes sign, which determines stability in a purely density-gradient-driven TEM. In the figure, $\omega_\lambda>0$ corresponds to instability (and associated \AE{}). It can be seen that this positive triangularity tokamak is unstable up to roughly $k^2=1/2$, and the magnetic well is relatively narrow \par 
The same information is displayed for a tokamak which has $\delta=-0.5$ in Fig. \ref{fig:neg-triangularity-tok}. It can be seen that the precession frequencies are unstable for a broader range of values for $k^2$. The \AE{} is furthermore weighted by the bounce-time of a particle, which can become very large at the bottom of a magnetic well in a negative triangularity tokamak. As such, the negative triangularity case (with the Miller vectors as chosen here) has \emph{higher} \AE{} than the positive triangularity case. \par  
We have tried various numerical experiments to assess the origin of this difference. From Eq. \eqref{eq:precession-freq-general}, we note that the term involving $1 - \lambda \hat{B} \propto v_\parallel^2$, and hence we identify this term as the curvature component of the drift. The term involving $\lambda \hat{B} \propto v_\perp^2$ on the other hand we identify as the gradient drift. Setting the term involving the parallel velocities equal to zero results in the found trends inverting, showcasing that this drive plays an important part in determining the precession. The poloidal curvature, $R_c^{-1}$, furthermore plays an important part. By setting this term equal to one in Eq. \eqref{eq:precession-freq-general}, we also find that negative triangularity is preferred over positive triangularity. Therefore, we postulate that this curvature drift plays an important part in determining stability. Importantly, the particles that experience curvature drive in negative triangularity tokamaks are the deeply trapped particles, which tend to be most unstable against the TEM with a density gradient. This is in contrast to positive triangularity tokamaks, where the most shallowly trapped particles experience significant curvature drive. These shallowly trapped particles however, are stabilised by the fact that they experience an averaged drift, and as such the curvature drive here is less deleterious.
\bibliographystyle{jpp}

\bibliography{references}

\begin{thebibliography}{55}
\expandafter\ifx\csname natexlab\endcsname\relax\def\natexlab#1{#1}\fi
\def\au#1{#1} \def\ed#1{#1} \def\yr#1{#1}\def\at#1{#1}\def\jt#1{\textit{#1}}
  \def\bt#1{#1}\def\bvol#1{\textbf{#1}} \def\vol#1{#1} \def\pg#1{#1}
  \def\publ#1{#1}\def\arxiv#1{#1}\def\org#1{#1}\def\st#1{\textit{#1}}

\bibitem[Balestri {\em et~al.\/}(2023)Balestri, Ball \&
  Coda]{balestri2023aspectplasmas}
{\sc \au{Balestri, A}, \au{Ball, J} \& \au{Coda, S}} \yr{2023}  \at{The aspect
  ratio dependence on confinement enhancement in negative triangularity plasmas
  (in preparation)} .

\bibitem[{C. Mercier} \& {N. Luc}(1974)]{MercierLuc1974}
{\sc \au{{C. Mercier}} \& \au{{N. Luc}}} \yr{1974}  \bt{{Report No. EUR-5127e
  140 (Commission of the European Communities, Brussels, 1974)}}.  \org{{\em
  Tech. Rep.\/}}.

\bibitem[Candy(2009)]{candy2009unified}
{\sc \au{Candy, J}} \yr{2009}  \at{A unified method for operator evaluation in
  local grad--shafranov plasma equilibria}.  \jt{Plasma Physics and Controlled
  Fusion}  \bvol{51}~(10),  \pg{105009}.

\bibitem[Connor {\em et~al.\/}(1983)Connor, Hastie \&
  Martin]{Connor1983EffectTokamak}
{\sc \au{Connor, J~W}, \au{Hastie, R~J} \& \au{Martin, T~J}} \yr{1983}
  \at{{Effect of pressure gradients on the bounce-averaged particle drifts in a
  tokamak}}.  \jt{Nucl. Fusion}  \bvol{23}~(12),  \pg{1702}.

\bibitem[Coppi \& Pegoraro(1977)]{coppi1977theory}
{\sc \au{Coppi, B} \& \au{Pegoraro, F}} \yr{1977}  \at{Theory of the ubiquitous
  mode}.  \jt{Nuclear Fusion}  \bvol{17}~(5),  \pg{969}.

\bibitem[Costello {\em et~al.\/}(2023)Costello, Proll, Plunk, Pueschel \&
  Alcus{\'o}n]{costello2023universal}
{\sc \au{Costello, P}, \au{Proll, JHE}, \au{Plunk, GG}, \au{Pueschel, MJ} \&
  \au{Alcus{\'o}n, JA}} \yr{2023}  \at{The universal instability in optimised
  stellarators}.  \jt{Journal of Plasma Physics}  \bvol{89}~(4),
  \pg{905890402}.

\bibitem[Dagazian \& Paris(1982)]{Dagazian1982TheBallooning}
{\sc \au{Dagazian, R~Y} \& \au{Paris, R~B}} \yr{1982}  \at{{The effects of high
  shear on ideal ballooning}}.  \jt{Plasma Physics}  \bvol{24}~(6),
  \pg{661--670}.

\bibitem[Dimits {\em et~al.\/}(2000)Dimits, Bateman, Beer, Cohen, Dorland,
  Hammett, Kim, Kinsey, Kotschenreuther, Kritz, Lao, Mandrekas, Nevins, Parker,
  Redd, Shumaker, Sydora \& Weiland]{Dimits2000ComparisonsSimulations}
{\sc \au{Dimits, A~M}, \au{Bateman, G}, \au{Beer, M~A}, \au{Cohen, B~I},
  \au{Dorland, W}, \au{Hammett, G~W}, \au{Kim, C}, \au{Kinsey, J~E},
  \au{Kotschenreuther, M}, \au{Kritz, A~H}, \au{Lao, L~L}, \au{Mandrekas, J},
  \au{Nevins, W~M}, \au{Parker, S~E}, \au{Redd, A~J}, \au{Shumaker, D~E},
  \au{Sydora, R} \& \au{Weiland, J}} \yr{2000}  \at{{Comparisons and physics
  basis of tokamak transport models and turbulence simulations}}.  \jt{Phys.
  Plasmas}  \bvol{7}~(3),  \pg{969--983}.

\bibitem[Duff {\em et~al.\/}(2022)Duff, Faber, Hegna, Pueschel \&
  Terry]{duff2022effect}
{\sc \au{Duff, JM}, \au{Faber, BJ}, \au{Hegna, CC}, \au{Pueschel, MJ} \&
  \au{Terry, PW}} \yr{2022}  \at{Effect of triangularity on
  ion-temperature-gradient-driven turbulence}.  \jt{Physics of Plasmas}
  \bvol{29}~(1).

\bibitem[Endres {\em et~al.\/}(2018)Endres, Sandrock \&
  Focke]{Endres2018AOptimisation}
{\sc \au{Endres, Stefan~C}, \au{Sandrock, Carl} \& \au{Focke, Walter~W}}
  \yr{2018}  \at{{A simplicial homology algorithm for Lipschitz optimisation}}.
   \jt{Journal of Global Optimization}  \bvol{72}~(2),  \pg{181--217}.

\bibitem[Gardner(1963)]{Gardner1963BoundPlasma}
{\sc \au{Gardner, Clifford~S}} \yr{1963}  \at{{Bound on the energy available
  from a plasma}}.  \jt{Phys. Fluids}  \bvol{6}~(6),  \pg{839--840}.

\bibitem[Hameiri(1983)]{Hameiri1983ThePlasmas}
{\sc \au{Hameiri, Eliezer}} \yr{1983}  \at{{The equilibrium and stability of
  rotating plasmas}}.  \jt{The Physics of Fluids}  \bvol{26}~(1),
  \pg{230--237}.

\bibitem[Hatch {\em et~al.\/}(2011)Hatch, Terry, Jenko, Merz, Pueschel, Nevins
  \& Wang]{hatch2011role}
{\sc \au{Hatch, DR}, \au{Terry, PW}, \au{Jenko, F}, \au{Merz, F}, \au{Pueschel,
  MJ}, \au{Nevins, WM} \& \au{Wang, E}} \yr{2011}  \at{Role of subdominant
  stable modes in plasma microturbulence}.  \jt{Physics of Plasmas}
  \bvol{18}~(5).

\bibitem[Helander(2017)]{Helander2017AvailablePlasmas}
{\sc \au{Helander, Per}} \yr{2017}  \at{{Available energy and ground states of
  collisionless plasmas}}.  \jt{J. Plasma Phys.}  \bvol{83}~(4).

\bibitem[Helander(2020)]{Helander2020AvailablePlasmas}
{\sc \au{Helander, Per}} \yr{2020}  \at{{Available energy of magnetically
  confined plasmas}}.  \jt{J. Plasma Phys.}  \bvol{86}~(2).

\bibitem[Helander \& Plunk(2015)]{helander2015universal}
{\sc \au{Helander, P} \& \au{Plunk, GG}} \yr{2015}  \at{The universal
  instability in general geometry}.  \jt{Physics of Plasmas}  \bvol{22}~(9).

\bibitem[Helander \& Sigmar(2005)]{Helander2005CollisionalPlasmas}
{\sc \au{Helander, Per} \& \au{Sigmar, Dieter~J}} \yr{2005} {\em {Collisional
  transport in magnetized plasmas}\/}.  \publ{Cambridge university press}.

\bibitem[Jenko {\em et~al.\/}(2001)Jenko, Dorland \&
  Hammett]{jenko2001critical}
{\sc \au{Jenko, F}, \au{Dorland, W} \& \au{Hammett, GW}} \yr{2001}
  \at{Critical gradient formula for toroidal electron temperature gradient
  modes}.  \jt{Physics of Plasmas}  \bvol{8}~(9),  \pg{4096--4104}.

\bibitem[Kesner {\em et~al.\/}(1995)Kesner, Ramos \&
  Gang]{Kesner1995CometTokamaks}
{\sc \au{Kesner, J}, \au{Ramos, J~J} \& \au{Gang, F~Y.}} \yr{1995}  \at{{Comet
  cross-section tokamaks}}.  \jt{Journal of Fusion Energy}  \bvol{14}~(4),
  \pg{361--371}.

\bibitem[Kessel {\em et~al.\/}(1994)Kessel, Manickam, Rewoldt \&
  Tang]{Kessel1994ImprovedShear}
{\sc \au{Kessel, C}, \au{Manickam, Jf}, \au{Rewoldt, G} \& \au{Tang, W~M}}
  \yr{1994}  \at{{Improved plasma performance in tokamaks with negative
  magnetic shear}}.  \jt{Phys. Rev. Lett.}  \bvol{72}~(8),  \pg{1212}.

\bibitem[Kinsey {\em et~al.\/}(2006)Kinsey, Waltz \&
  Candy]{Kinsey2006TheSimulations}
{\sc \au{Kinsey, J~E}, \au{Waltz, R~E} \& \au{Candy, J}} \yr{2006}  \at{{The
  effect of safety factor and magnetic shear on turbulent transport in
  nonlinear gyrokinetic simulations}}.  \jt{Phys. Plasmas}  \bvol{13}~(2),
  \pg{022305}.

\bibitem[Kolmes \& Fisch(2022)]{kolmes2022minimum}
{\sc \au{Kolmes, EJ} \& \au{Fisch, NJ}} \yr{2022}  \at{Minimum stabilizing
  energy release for mixing processes}.  \jt{Physical Review E}
  \bvol{106}~(5),  \pg{055209}.

\bibitem[Kolmes \& Fisch(2020)]{Kolmes2020RecoveringOperations}
{\sc \au{Kolmes, E~J} \& \au{Fisch, N~J}} \yr{2020}  \at{{Recovering Gardner
  restacking with purely diffusive operations}}.  \jt{Phys. Rev. E}
  \bvol{102}~(6),  \pg{63209}.

\bibitem[Kolmes {\em et~al.\/}(2020)Kolmes, Helander \&
  Fisch]{Kolmes2020AvailableRearrangements}
{\sc \au{Kolmes, E~J}, \au{Helander, P} \& \au{Fisch, N~J}} \yr{2020}
  \at{{Available energy from diffusive and reversible phase space
  rearrangements}}.  \jt{Phys. Plasmas}  \bvol{27}~(6),  \pg{062110}.

\bibitem[Landreman {\em et~al.\/}(2015)Landreman, Antonsen~Jr \&
  Dorland]{landreman2015universal}
{\sc \au{Landreman, Matt}, \au{Antonsen~Jr, Thomas~M} \& \au{Dorland, William}}
  \yr{2015}  \at{Universal instability for wavelengths below the ion larmor
  scale}.  \jt{Physical review letters}  \bvol{114}~(9),  \pg{095003}.

\bibitem[Lang {\em et~al.\/}(2008)Lang, Parker \& Chen]{lang2008nonlinear}
{\sc \au{Lang, Jianying}, \au{Parker, Scott~E} \& \au{Chen, Yang}} \yr{2008}
  \at{Nonlinear saturation of collisionless trapped electron mode turbulence:
  Zonal flows and zonal density}.  \jt{Physics of Plasmas}  \bvol{15}~(5).

\bibitem[Mackenbach {\em et~al.\/}(2023{\natexlab{{\em a\/}}})Mackenbach,
  Proll, Wakelkamp \& Helander]{Mackenbach2023AvailableTransport}
{\sc \au{Mackenbach, R.J.J.}, \au{Proll, J.H.E.}, \au{Wakelkamp, R.} \&
  \au{Helander, P.}} \yr{2023{\natexlab{{\em a\/}}}}  \at{The available energy
  of trapped electrons: a nonlinear measure for turbulent transport}.
  \jt{Journal of Plasma Physics}  \bvol{89}~(5),  \pg{905890513}.

\bibitem[Mackenbach {\em et~al.\/}(2023{\natexlab{{\em b\/}}})Mackenbach, Duff,
  Gerard, Proll, Helander \& Hegna]{mackenbach2023drift}
{\sc \au{Mackenbach, R. J.~J.}, \au{Duff, J.~M.}, \au{Gerard, M.~J.},
  \au{Proll, J. H.~E.}, \au{Helander, P.} \& \au{Hegna, C.~C.}}
  \yr{2023{\natexlab{{\em b\/}}}}  \at{{Bounce-averaged drifts: Equivalent
  definitions, numerical implementations, and example cases}}.  \jt{Physics of
  Plasmas}  \bvol{30}~(9),  \pg{093901}.

\bibitem[Mackenbach {\em et~al.\/}(2022)Mackenbach, Proll \&
  Helander]{Mackenbach2022AvailableTransport}
{\sc \au{Mackenbach, R J~J}, \au{Proll, Josefine H~E} \& \au{Helander, P}}
  \yr{2022}  \at{{Available Energy of Trapped Electrons and Its Relation to
  Turbulent Transport}}.  \jt{Physical Review Letters}  \bvol{128}~(17),
  \pg{175001}.

\bibitem[Marinoni {\em et~al.\/}(2019)Marinoni, Austin, Hyatt, Walker, Candy,
  Chrystal, Lasnier, McKee, Odstr{\v{c}}il, Petty, Porkolab, Rost, Sauter,
  Smith, Staebler, Sung, Thome, Turnbull \& Zeng]{Marinoni2019H-modeDIII-D}
{\sc \au{Marinoni, A}, \au{Austin, M~E}, \au{Hyatt, A~W}, \au{Walker, M~L},
  \au{Candy, J}, \au{Chrystal, C}, \au{Lasnier, C~J}, \au{McKee, G~R},
  \au{Odstr{\v{c}}il, T}, \au{Petty, C~C}, \au{Porkolab, M}, \au{Rost, J~C},
  \au{Sauter, O}, \au{Smith, S~P}, \au{Staebler, G~M}, \au{Sung, C}, \au{Thome,
  K~E}, \au{Turnbull, A~D} \& \au{Zeng, L}} \yr{2019}  \at{{H-mode grade
  confinement in L-mode edge plasmas at negative triangularity on DIII-D}}.
  \jt{Physics of Plasmas}  \bvol{26}~(4),  \pg{042515}.

\bibitem[Merlo {\em et~al.\/}(2015)Merlo, Brunner, Sauter, Camenen,
  G{\"{o}}rler, Jenko, Marinoni, Told \&
  Villard]{Merlo2015InvestigatingTransport}
{\sc \au{Merlo, G}, \au{Brunner, S}, \au{Sauter, Olivier}, \au{Camenen, Y},
  \au{G{\"{o}}rler, T}, \au{Jenko, F}, \au{Marinoni, A}, \au{Told, D} \&
  \au{Villard, Laurent}} \yr{2015}  \at{{Investigating profile stiffness and
  critical gradients in shaped TCV discharges using local gyrokinetic
  simulations of turbulent transport}}.  \jt{Plasma Phys. Control. Fusion}
  \bvol{57}~(5),  \pg{054010}.

\bibitem[Merlo {\em et~al.\/}(2019)Merlo, Fontana, Coda, Hatch, Janhunen, Porte
  \& Jenko]{merlo2019turbulent}
{\sc \au{Merlo, G}, \au{Fontana, Matteo}, \au{Coda, Stephano}, \au{Hatch, D},
  \au{Janhunen, S}, \au{Porte, Laurie} \& \au{Jenko, F}} \yr{2019}
  \at{{Turbulent transport in TCV plasmas with positive and negative
  triangularity}}.  \jt{Physics of Plasmas}  \bvol{26}~(10),  \pg{102302}.

\bibitem[Merlo {\em et~al.\/}(2021)Merlo, Huang, Marini, Brunner, Coda, Hatch,
  Jarema, Jenko, Sauter \& Villard]{Merlo2021NonlocalPlasmas}
{\sc \au{Merlo, G}, \au{Huang, Z}, \au{Marini, C}, \au{Brunner, S}, \au{Coda,
  S}, \au{Hatch, D}, \au{Jarema, D}, \au{Jenko, F}, \au{Sauter, O} \&
  \au{Villard, L}} \yr{2021}  \at{{Nonlocal effects in negative triangularity
  TCV plasmas}}.  \jt{Plasma Phys. Control. Fusion}  \bvol{63}~(4),
  \pg{044001}.

\bibitem[Merlo \& Jenko(2023)]{merlo2023interplay}
{\sc \au{Merlo, Gabriele} \& \au{Jenko, Frank}} \yr{2023}  \at{Interplay
  between magnetic shear and triangularity in ion temperature gradient and
  trapped electron mode dominated plasmas}.  \jt{Journal of Plasma Physics}
  \bvol{89}~(1),  \pg{905890104}.

\bibitem[Miller {\em et~al.\/}(1989)Miller, Chu, Dominguez \&
  Ohkawa]{Miller1989MaximumShaping}
{\sc \au{Miller, R~L}, \au{Chu, M~S}, \au{Dominguez, R~R} \& \au{Ohkawa, T}}
  \yr{1989}  \at{{Maximum J tokamak by plasma shaping}}.  \jt{Comments on
  Plasma Physics and Controlled Fusion}  \bvol{12}~(3),  \pg{125--132}.

\bibitem[Miller {\em et~al.\/}(1998)Miller, Chu, Greene, Lin-Liu \&
  Waltz]{Miller1998NoncircularModel}
{\sc \au{Miller, R~L}, \au{Chu, Ming-Sheng}, \au{Greene, J~M}, \au{Lin-Liu,
  Y~R} \& \au{Waltz, R~E}} \yr{1998}  \at{{Noncircular, finite aspect ratio,
  local equilibrium model}}.  \jt{Physics of Plasmas}  \bvol{5}~(4),
  \pg{973--978}.

\bibitem[Miller {\em et~al.\/}(1995)Miller, Waelbroeck, Hassam \&
  Waltz]{Miller1995StabilizationRotation}
{\sc \au{Miller, R~L}, \au{Waelbroeck, F~L}, \au{Hassam, A~B} \& \au{Waltz,
  R~E}} \yr{1995}  \at{{Stabilization of ballooning modes with sheared toroidal
  rotation}}.  \jt{Physics of Plasmas}  \bvol{2}~(10),  \pg{3676--3684}.

\bibitem[Miyamoto(2005)]{miyamoto2005plasma}
{\sc \au{Miyamoto, Kenro}} \yr{2005} {\em Plasma physics and controlled nuclear
  fusion\/}, ,  \vol{vol.~38}.  \publ{Springer Science \& Business Media}.

\bibitem[Nelson {\em et~al.\/}(2022)Nelson, Paz-Soldan \&
  Saarelma]{nelson2022prospects}
{\sc \au{Nelson, AO}, \au{Paz-Soldan, C} \& \au{Saarelma, S}} \yr{2022}
  \at{{Prospects for H-mode inhibition in negative triangularity tokamak
  reactor plasmas}}.  \jt{Nuclear Fusion}  \bvol{62}~(9),  \pg{096020}.

\bibitem[Proll {\em et~al.\/}(2022)Proll, Plunk, Faber, G\"{o}rler, Helander,
  McKinney, Pueschel, Smith \& Xanthopoulos]{Proll2022mitigation}
{\sc \au{Proll, J.H.E.}, \au{Plunk, G.G.}, \au{Faber, B.J.}, \au{G\"{o}rler,
  T.}, \au{Helander, P.}, \au{McKinney, I.J.}, \au{Pueschel, M.J.}, \au{Smith,
  H.M.} \& \au{Xanthopoulos, P.}} \yr{2022}  \at{{Turbulence mitigation in
  maximum-J stellarators with electron-density gradient}}.  \jt{Journal of
  Plasma Physics}  \bvol{88}~(1),  \pg{905880112}.

\bibitem[Proll {\em et~al.\/}(2012)Proll, Helander, Connor \&
  Plunk]{proll2012resilience}
{\sc \au{Proll, Josefine Henriette~Elise}, \au{Helander, Per}, \au{Connor,
  John~William} \& \au{Plunk, GG}} \yr{2012}  \at{Resilience of
  quasi-isodynamic stellarators against trapped-particle instabilities}.
  \jt{Physical Review Letters}  \bvol{108}~(24),  \pg{245002}.

\bibitem[Pueschel {\em et~al.\/}(2016)Pueschel, Faber, Citrin, Hegna, Terry \&
  Hatch]{pueschel2016stellarator}
{\sc \au{Pueschel, MJ}, \au{Faber, BJ}, \au{Citrin, J}, \au{Hegna, CC},
  \au{Terry, PW} \& \au{Hatch, DR}} \yr{2016}  \at{Stellarator turbulence:
  subdominant eigenmodes and quasilinear modeling}.  \jt{Physical review
  letters}  \bvol{116}~(8),  \pg{085001}.

\bibitem[Rettig {\em et~al.\/}(1997)Rettig, Peebles, Doyle, Burrell,
  Greenfield, Staebler \& Rice]{Rettig1997MicroturbulenceDischarges}
{\sc \au{Rettig, C~L}, \au{Peebles, W~A}, \au{Doyle, E~J}, \au{Burrell, K~H},
  \au{Greenfield, C}, \au{Staebler, G~M} \& \au{Rice, B~W}} \yr{1997}
  \at{{Microturbulence reduction during negative central shear tokamak
  discharges}}.  \jt{Phys. Plasmas}  \bvol{4}~(11),  \pg{4009--4016}.

\bibitem[Roach {\em et~al.\/}(1995)Roach, Connor \&
  Janjua]{Roach1995TrappedTokamaks}
{\sc \au{Roach, C~M}, \au{Connor, J~W} \& \au{Janjua, S}} \yr{1995}
  \at{{Trapped particle precession in advanced tokamaks}}.  \jt{Plasma Phys.
  Control. Fusion}  \bvol{37}~(6),  \pg{679}.

\bibitem[Rodrigues \& Coroado(2018)]{Rodrigues2018LocalPlasmas}
{\sc \au{Rodrigues, Paulo} \& \au{Coroado, André}} \yr{2018}  \at{{Local
  updown asymmetrically shaped equilibrium model for tokamak plasmas}}.
  \jt{Nuclear Fusion}  \bvol{58}~(10),  \pg{106040}.

\bibitem[Rodriguez \& Mackenbach(2023)]{rodriguez2023trapped}
{\sc \au{Rodriguez, E} \& \au{Mackenbach, RJJ}} \yr{2023}  \at{Trapped-particle
  precession and modes in quasi-symmetric stellarators and tokamaks: a
  near-axis perspective}.  \jt{arXiv preprint arXiv:2308.00960} .

\bibitem[Romanelli(1989)]{romanelli1989ion}
{\sc \au{Romanelli, Ft}} \yr{1989}  \at{Ion temperature-gradient-driven modes
  and anomalous ion transport in tokamaks}.  \jt{Physics of Fluids B: Plasma
  Physics}  \bvol{1}~(5),  \pg{1018--1025}.

\bibitem[Rosenbluth \& Sloan(1971)]{rosenbluth1971finite}
{\sc \au{Rosenbluth, M} \& \au{Sloan, ML}} \yr{1971}  \at{Finite-$\beta$
  stabilization of the collisionless trapped particle instability}.  \jt{The
  Physics of Fluids}  \bvol{14}~(8),  \pg{1725--1741}.

\bibitem[Saarelma {\em et~al.\/}(2021)Saarelma, Austin, Knolker, Marinoni,
  Paz-Soldan, Schmitz \& Snyder]{Saarelma2021BallooningTokamak}
{\sc \au{Saarelma, Samuli}, \au{Austin, Max~E}, \au{Knolker, M}, \au{Marinoni,
  Alessandro}, \au{Paz-Soldan, Carlos}, \au{Schmitz, Lothar} \& \au{Snyder,
  Philip~B}} \yr{2021}  \at{{Ballooning instability preventing the H-mode
  access in plasmas with negative triangularity shape on the DIII–D
  tokamak}}.  \jt{Plasma Physics and Controlled Fusion}  \bvol{63}~(10),
  \pg{105006}.

\bibitem[Staebler \& Kinsey(2010)]{staebler2010electron}
{\sc \au{Staebler, GM} \& \au{Kinsey, JE}} \yr{2010}  \at{Electron collisions
  in the trapped gyro-landau fluid transport model}.  \jt{Physics of Plasmas}
  \bvol{17}~(12),  \pg{122309}.

\bibitem[Staebler {\em et~al.\/}(2007)Staebler, Kinsey \&
  Waltz]{staebler2007theory}
{\sc \au{Staebler, GM}, \au{Kinsey, JE} \& \au{Waltz, RE}} \yr{2007}  \at{A
  theory-based transport model with comprehensive physics}.  \jt{Physics of
  Plasmas}  \bvol{14}~(5),  \pg{055909}.

\bibitem[Staebler {\em et~al.\/}(2021)Staebler, Belli, Candy, Kinsey, Dudding
  \& Patel]{staebler2021verification}
{\sc \au{Staebler, Gary~M}, \au{Belli, EA}, \au{Candy, J}, \au{Kinsey, JE},
  \au{Dudding, H} \& \au{Patel, B}} \yr{2021}  \at{Verification of a
  quasi-linear model for gyrokinetic turbulent transport}.  \jt{Nuclear Fusion}
   \bvol{61}~(11),  \pg{116007}.

\bibitem[Staebler {\em et~al.\/}(2020)Staebler, Candy, Belli, Kinsey, Bonanomi
  \& Patel]{staebler2020geometry}
{\sc \au{Staebler, Gary~M}, \au{Candy, Jeffrey}, \au{Belli, Emily~A},
  \au{Kinsey, Jon~E}, \au{Bonanomi, N} \& \au{Patel, Bhavin}} \yr{2020}
  \at{Geometry dependence of the fluctuation intensity in gyrokinetic
  turbulence}.  \jt{Plasma Physics and Controlled Fusion}  \bvol{63}~(1),
  \pg{015013}.

\bibitem[Strait {\em et~al.\/}(1997)Strait, Casper, Chu, Ferron, Garofalo,
  Greenfield, La~Haye, Lao, Lazarus \& Miller]{Strait1997StabilityTokamak}
{\sc \au{Strait, E~J}, \au{Casper, T~A}, \au{Chu, M~S}, \au{Ferron, J~R},
  \au{Garofalo, A}, \au{Greenfield, C~M}, \au{La~Haye, R~J}, \au{Lao, L~L},
  \au{Lazarus, E~A} \& \au{Miller, R~L}} \yr{1997}  \at{{Stability of negative
  central magnetic shear discharges in the DIII-D tokamak}}.  \jt{Phys.
  Plasmas}  \bvol{4}~(5),  \pg{1783--1791}.

\bibitem[Turnbull {\em et~al.\/}(1999)Turnbull, Lin-Liu, Miller, Taylor \&
  Todd]{Turnbull1999ImprovedTokamaks}
{\sc \au{Turnbull, A~D}, \au{Lin-Liu, Y~R}, \au{Miller, R~L}, \au{Taylor, T~S}
  \& \au{Todd, T~N}} \yr{1999}  \at{{Improved magnetohydrodynamic stability
  through optimization of higher order moments in cross-section shape of
  tokamaks}}.  \jt{Physics of Plasmas}  \bvol{6}~(4),  \pg{1113--1116}.

\end{thebibliography}

\end{document}